\documentclass[12pt,english]{revtex4-1}
\usepackage{times}
\usepackage{array}
\usepackage{longtable}
\usepackage{float}
\usepackage{amsmath}
\usepackage{comment}
\usepackage{bm}
\usepackage[normalem]{ulem} 
\usepackage{graphicx}
\usepackage[FIGTOPCAP]{subfigure}
\usepackage{amssymb}
\usepackage{color}

\makeatletter
\usepackage{fancyhdr}
\usepackage{babel}
\makeatother

\begin{document}

\title{A Deep Learning Approach to Describing the Plasma Sheath}

\author{E. Webb}

\address{Nuclear Engineering Program, Department of Materials Science and Engineering, University of Florida, Gainesville, FL 32611, United States of America}

\author{Yuzhi Li}

\address{Nuclear Engineering Program, Department of Materials Science and Engineering, University of Florida, Gainesville, FL 32611, United States of America}

\author{Christopher J. McDevitt}

\address{Nuclear Engineering Program, Department of Materials Science and Engineering, University of Florida, Gainesville, FL 32611, United States of America}

\date{\today}

\begin{abstract}

Despite their ubiquity, the rich physics present in a plasma sheath has inhibited the development of a generally applicable description of this critical region. The present study utilizes a physics-informed neural network (PINN) to evaluate a hierarchy of models of the plasma sheath. Unlike traditional deep learning methods, PINNs use the governing PDEs to constrain the predictions of a neural network, and thus do not require any experimental or simulation data to train. In this work, we utilize a PINN to identify the parametric solution to fluid models of different physics fidelity of the plasma sheath. While the offline training time of the PINN is often longer than a traditional solver, once trained, the PINN is able to efficiently predict the sheath profiles across a broad range of parameter regimes, thus yielding an effective surrogate of the plasma sheath. 

\end{abstract}
\maketitle

\newpage

\section{Introduction}

The plasma-sheath transition strongly influences a diverse range of applications including magnetic fusion \cite{stangeby}, electric propulsion \cite{kaganovich2020physics}, plasma diagnostics \cite{hutchinson2002principles}, and plasma processing \cite{lieberman1994principles}. Despite extensive research and the attention obtained from a broad range of applications, this region still presents significant challenges for modeling and simulation to overcome. This is largely due to the rich physics present in the sheath, where fundamental quantities, such as the Bohm criterion, are strongly impacted by a diverse range of plasma phenomena, including collisions \cite{godyak1982, riemann1997collisions}, an oblique-angle magnetic field \cite{chodura1982plasma}, and plasma instabilities \cite{hershkowitz2005sheaths, baalrud2009instability}. The present work seeks to demonstrate the potential of physics-constrained machine learning methods~\cite{karniadakis2021physics} to provide a novel path through which this classic problem can be treated. 

This is accomplished by considering a hierarchy of fluid models of the plasma sheath, where it is demonstrated that a physics-informed neural network (PINN)~\cite{Raissi,van1995neural} is able to provide an accurate solution of the sheath profile across a broad range of plasma conditions. Specifically, unlike the majority of machine learning (ML) techniques that have, to date, been applied to low-temperature plasmas \cite{trieschmann2023machine}, PINNs seek to embed physical constraints into the training of a neural network. Broadly, imposing physical constraints during the training of a neural network serves three purposes. For the case of sparse datasets~\cite{ahn2025deep}, physical constraints can be used to regularize predictions of the neural network, thus providing a robust mechanism for interpolating across regions that would otherwise be left largely unconstrained. Neural Networks have a tendency to overfit data, especially as the sparsity increases; however, using the physical equations to penalize the network reduces overfitting. For the case where the physics of the system is completely specified, including both the governing equations along with initial and boundary conditions, this approach can be used to fully constrain the system even in the absence of data, such that the PINN can be used as an ordinary differential equation (ODE) or partial differential equation (PDE) solver. In contrast, for cases where a large dataset is present, but the physics of the problem is incompletely understood, the same PINN framework can be employed to solve an inverse problem, where synthetic or experimental data are used to close the incomplete set of imposed physics~\cite{Raissi, mathews2022deep}, an approach often referred to as physics discovery \cite{brunton2016,rudy2017,alves2022}.

In this paper, we will assume the limit of no synthetic or experimental data, and develop a series of PINNs that solve the system of ODEs characterizing a one-dimensional sheath. For this approach, PINNs solve an optimization problem, where the PINN seeks to minimize the loss associated with the residuals of a system of ODEs together with boundary conditions. If successful, this results in the identification of a solution to the sheath profiles. This approach is applied to several fluid models of the sheath, some of which can be straightforwardly solved with an ODE solver such as Runge-Kutta~\cite{beving2022sheath,riemann2005plasma,robertson2013sheaths, holmes2007introduction}, but others that require a more involved approach to obtain a steady state solution~\cite{alvarez2020plasma, colella1999conservative}. A distinguishing feature of the PINN approach is that while the offline training time of the PINN is often longer than a traditional ODE or PDE solver, a single PINN is able to learn the parametric solution to the ODE or PDE system of interest \cite{sun2020surrogate, mcdevitt2024physics, jang2024grad, osborne2026physics}. Noting that the online inference time of a PINN is typically on the order of microseconds per prediction, once trained, a PINN can be used to predict sheath profiles across a broad range of physics parameters. Physics-informed neural networks thus provide a means through which efficient surrogates for the plasma sheath can be constructed. In this paper, we build three surrogates of the plasma sheath, trained on fluid models of varying physics fidelity. The resulting models provide an efficient means of investigating fundamental quantities such as the Bohm criterion, sheath width, and potential drop and how they depend on parameters such as the ion-neutral mean-free-path, ion-electron temperature ratio, and ion mass, as well as demonstrate the ability of PINNs to accurately treat sheath physics.

The remainder of this paper is organized as follows. The physics-informed machine learning approach employed is briefly described in Sec. \ref{sec:PIML}. In Sec. \ref{sec:FM}, three fluid models of the plasma sheath with increasing physics fidelity are defined and the results are included in each section. Section \ref{sec:C} describes conclusions and future directions.

\section{\label{sec:PIML}Physics-Constrained Machine Learning}

Neural networks (NN) are powerful tools that are well known for their ability to identify patterns in data \cite{bishop2006pattern}. Data-driven approaches are successful when making predictions within the scope of their training data, but require sufficient data to converge well. When data is scarce or expensive to obtain, a data-driven approach becomes prohibitive. Due to the difficulty of obtaining plasma sheath data, we find that physics-constrained approaches \cite{karniadakis2021physics} provide a path to overcome this relative scarcity of data. Here, physical constraints are embedded in the training of a neural network (NN), providing a natural regularization term and mechanism for extrapolating beyond available data.
This can be accomplished by the identification of NN architectures that enforce select properties of the solution~\cite{burby2020fast}, or by enforcing physical properties as soft constraints in the loss function. Regarding the latter, PINNs \cite{Raissi, cuomo2022, wang2023expert, toscano2025pinns} include the residuals, which is simply an ODE or PDE rearranged to equal zero, of the governing equations into the loss function along with the data, yielding a loss function of the form:
\begin{subequations}
\begin{equation}
    \label{eq:31} 
    \text{Loss} = L_{Data} + L_{ODE},
\end{equation}
\begin{equation}
    \label{eq:31sub1} 
    L_{Data} = \frac{1}{N_{Data}} \sum^{N_{Data}}_i \left[ \phi \left( x_i;\bm{a}_i\right) - \phi_i\right]^2,
\end{equation}
\begin{equation}
    \label{eq:31sub2} 
    L_{ODE} = \frac{w_{ODE}}{N_{ODE}} \sum^{N_{ODE}}_i \mathcal{R}^2 \left( x_i;\bm{a}_i\right),
\end{equation}
\end{subequations}
where $\phi$ is a field predicted by the PINN, $x_i$ is the spatial coordinate, $\bm{a}$ are the physical parameters, $\mathcal{R}$ is the residual of an ODE, and $w_{ODE}$ is the relative weight applied to the ODE component of the loss function. For a system of ODEs multiple data and ODE loss terms will be present, but for simplicity we only list one each here. The first term corresponds to losses against available data, which could include initial or boundary conditions, and the second term represents losses associated with the residual of an ODE. By minimizing the loss, a PINN seeks to find a solution that obeys both the available data set together with the governing equations. For the case where experimental or synthetic data are unavailable, but boundary and initial conditions are specified, the successful minimization of the loss would imply that the PINN has found a solution to ODE(s). 

\begin{figure}
\centering
\subfigure{\includegraphics[scale=0.5]{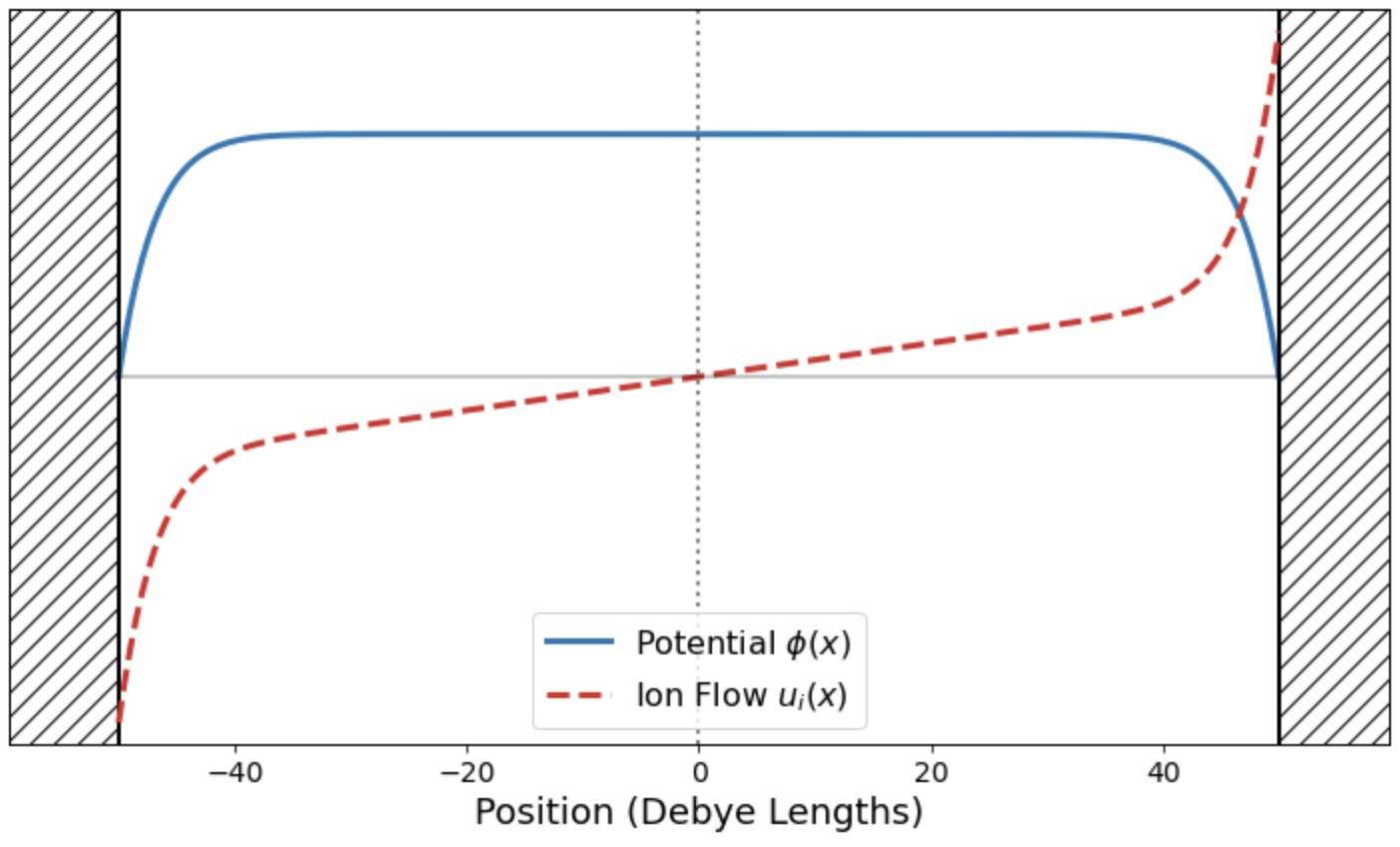}}
\caption{Schematic of the problem geometry. Absorbing walls are placed at $x=\pm L$ ($L=50$ in this example), resulting in even parity for the electrostatic potential, ion and electron densities, and ion and electron temperatures. Ion and electron flows and heat fluxes will have odd parity. Representative electrostatic potential and ion flow profiles are sketched as examples.}
\label{fig:21}
\end{figure}

This work will focus on the data-free limit, where we find that PINNs are able to rely entirely on the physical constraints without needing any data to guide the solution, provided the problem is well posed with adequate boundary conditions. Normally, boundary and initial conditions would be terms in the loss function as well. However, the framework for PINNs is flexible and allows for the manual addition of layers where boundary and initial conditions can be exactly enforced as hard constraints \cite{mcdevitt2024Navier}. This is beneficial to the optimizer for two reasons. First it removes terms from the loss function allowing the optimizer to focus more on the terms that are present, and second the boundary and initial conditions are precisely satisfied. In this way, boundary conditions are enforced a priori, guiding the network to a solution that obeys both the physics and the boundary conditions. As a simple example, the electrostatic potential $\phi$ can be forced to vanish at walls located at $x=\pm L$ (see Fig. \ref{fig:21} for the geometry employed in this work) by introducing the transform:
\begin{equation}
    \label{eq:32} 
    \phi = \frac{(L-x)(L+x)}{L^2}\phi_{NN} = \frac{L^2-x^2}{L^2} \phi_{NN},
\end{equation}
where $\phi_{NN}$ is the output of the hidden layers of the NN. Noting that $\phi$ will be even for the symmetric geometry employed, this property can be enforced by introducing a second layer, referred to as an input transform, that takes the input $x$ into the NN and squares it, such that $\phi_{NN} = \phi_{NN} \left( x^2\right)$. From this simple input transform, together with Eq. (\ref{eq:32}), it can be verified that $\phi \left( -x \right) = \phi \left( x\right)$, thus automatically satisfying the parity constraint. With the inclusion of these two additional layers, $\phi$ will automatically vanish at the walls, along with satisfy the Neumann condition $d\phi/dx = 0$ at $x=0$, such that these boundary conditions do not need to be included in the loss function. However, values without even parity are multiplied by an odd function in the output transform to avoid enforcing the same Neumann condition. For example, the output transform for velocity has a form like: $u=u_{NN}\left(x^2\right)\frac{x}{L}$ which both enforces a Dirichlet boundary at $x=0$ and preserves the expected odd symmetry of the flow. The precise enforcement of these symmetries also enable us to capture the sheath solution across the entire domain even when only explicitly solving one side [$x\in\left( 0, L\right)$, for example].

PINNs can be trained to provide predictions for the output fields as a function of several input parameters. Consequently, PINNs can be employed to explore large parameter spaces efficiently. Another feature of PINNs, or neural networks in general, is that no spatial discretization is required, since neural networks themselves are continuous and differentiable as long as a continuous activation function is used. As a result, the various derivatives that appear in PDEs can be computed by automatic differentiation, a feature that is standard in software packages like TensorFlow \cite{abadi2016tensorflow} and PyTorch \cite{paszke2019pytorch}. Additionally, since NNs are continuous and differentiable, predictions can be made at an arbitrary point within the trained domain. These predictions take typically a few microseconds each, and when combined with the parametric capabilities of PINNs, allow PINNs to serve as a rapid surrogate model. 

An important choice when designing PINNs is the choice of optimizer. Common optimizers used for machine learning include ADAM~\cite{kingma2014adam} and limited memory BFGS (L-BFGS) ~\cite{liu1989limited}. In recent work, Ref. \cite{kiyani2025optimizer} finds that second order optimization methods, such as Self-Scaled Broyden (SSBroyden), offer significant improvement over first order optimizers such as ADAM for physics based approaches. Furthermore, SSBroyden was shown to achieve similar levels of convergence for drastically smaller NNs \cite{urban2025unveiling}, freeing up memory for other parts of the training and reducing the computational cost. We have chosen to use ADAM through the initial training and SSBroyden to fine tune the solution. SSBroyden is found to drastically improve the convergence of the model, with the loss typically dropping by orders of magnitude during this second phase of training.

\section{\label{sec:FM}Fluid Models}

This section will describe the multi-fluid models of the plasma sheath employed in this work. The three models of the sheath considered incorporate an increasingly complex set of physics and are meant to demonstrate the flexibility of PINNs for treating the plasma sheath. For all three models, we will consider 1D geometry, with a plasma bounded by absorbing walls at $x \pm L$ (see Fig. \ref{fig:21}). For each fluid system, particle balance will be achieved by the introduction of a volumetric particle source which balances losses to the absorbing boundaries, thus allowing a steady state solution to be found. 

\subsection{\label{sec:CS}Constant Source}

\subsubsection{\label{sec:MCS}Model}
A minimal consistent model of the sheath can be formed by introducing a constant uniform source of ions and electrons throughout the plasma volume to balance losses to the surrounding walls (see Fig. \ref{fig:21}). The temperature is held constant for ions and electrons, under the assumption that the heat conductivity is sufficient to force a nearly constant temperature profile. The plasma will be assumed to be composed of ions and electrons together with a stationary and constant neutral background. Collisions with neutrals will be accounted for by including an ion-neutral drag whose strength is set by the ion-neutral mean-free-path $\lambda_{in}$. Under these assumptions, and with the addition of the Poisson equation to solve for the electric potential, a simple fluid model of the plasma sheath can be written as~\cite{stangeby}: 
\begin{subequations}
\label{eq:ConstSource}
\begin{equation}
    \label{eq:211} 
    \epsilon_0\frac{\partial^2\phi}{\partial x^2} = -e(n_i-n_e),
\end{equation}
\begin{equation}
    \label{eq:212} 
    \frac{\partial}{\partial x}(n_iu_i) = S_0,
\end{equation}
\begin{equation}
    \label{eq:213} 
    \frac{\partial}{\partial x}(n_eu_e) = S_0,
\end{equation}
\begin{equation}
    \label{eq:214} 
    T_e\frac{\partial n_e}{\partial x} = en_e\frac{\partial \phi}{\partial x},
\end{equation}
\begin{equation}
    \label{eq:215} 
    m_in_iu_i \frac{\partial u_i}{\partial x} + T_i\frac{\partial n_i}{\partial x} = -en_i\frac{\partial \phi}{\partial x} - m_iu_iS_0 + R_{in},
\end{equation}
\end{subequations}
where we took the ions to be singly ionized, included ion collisions with a stationary neutral population, $R_{in}$ is the ion-neutral collisional drag, $S_0$ is a constant particle source, and we have neglected electron inertia in the electron momentum equation. Furthermore, the continuity equations can be analytically integrated, yielding
\begin{equation}
    \label{eq:216} 
    n_su_s = xS_0,
\end{equation}
where the subscript $s$ indicates either ions or electrons. Additionally, the electron momentum equation can be integrated, giving Boltzmann electrons:
\begin{equation}
    \label{eq:217} 
    n_e = n_we^{e\phi/T_e},
\end{equation}
where $n_w$ is the electron density at the wall where the electric potential equals zero. The electron density at the wall is found by taking the electron velocity distribution at the wall to be a one-sided Maxwellian and integrating for all positive velocities. The 1D particle flux density into the wall is then $\Gamma_w=\frac{1}{4}n_wc_e \text{ m}^{-2}\text{s}^{-1}$~\cite{stangeby}, where $c_e = \left[8T_e/(\pi m_e)\right]^{1/2} \text{ m s}^{-1}$ is the average particle velocity. From Eq. (\ref{eq:216}), $\Gamma_w=S_0L_x$, thus $n_w = S_0L_x\sqrt{2\pi m_e/T_e} \text{ m}^{-3}$. Introducing the normalizations:

\begin{equation}
x \to \frac{x}{\lambda_{De}},\quad \phi \to \frac{e\phi}{T_e}, \quad u_s \to \frac{u_s}{C_s}, \quad n_s \to \frac{n_s}{n_0},
\label{eq:normalizations}
\end{equation}
where $\lambda_{De} \equiv \sqrt{\epsilon_0T_e / \left( n_0e^2 \right)}$ is the Debye length, $n_0 \equiv S_0 L_x / C_s$, $C_s \equiv \sqrt{T_e/m_i}$ is the sound speed, and $L_x$ is half the domain size in meters, and is normalized as $L = L_x / \lambda_{De}$. Written in residual form with these normalizations, Eq. (\ref{eq:ConstSource}) becomes:
\begin{subequations}
\label{eq:Model1}
\begin{equation}
    \label{eq:218} 
     0 = n_we^{\phi}-n_i - \frac{\partial^2\phi}{\partial x^2},
\end{equation}
\begin{equation}
    \label{eq:219} 
      0= -n_i\frac{\partial \phi}{\partial x} - \frac{T_i}{T_e}\frac{\partial n_i}{\partial x} - \frac{u_i}{L} - n_iu_i\left(\frac{\lambda_{De}}{\lambda_{in}}\right) - n_iu_i \frac{\partial u_i}{\partial x},
\end{equation}
\end{subequations}
where $\phi$ and $n_i$ are the dependent variables, or outputs of the model, and $u_i = x/(Ln_i)$. The PINN optimizes these equations by adjusting the outputs so the right hand sides (RHS) of the equations are close to zero. After applying normalizations, $\Gamma_w=1$, $n_w = \sqrt{2\pi m_e/m_i}$, and $R_{in}$ can be parameterized as $n_iu_i(\lambda_{De}/\lambda_{in}),$ with $\lambda_{in}$ the ion-neutral mean free path. The ratio, $\lambda_{De}/\lambda_{in}$, indicates the collisionality of the plasma. 

To fully specify the sheath solution we will need to impose boundary conditions on the system. Applying symmetry about $x=0$ to the domain allows us to only solve for one side of the domain. We will choose the positive side, yielding a computational domain from $x=0$ to $x=L$. Due to the symmetry of the solution, we can apply the boundary conditions $u_i=u_e=0$ and $\partial n_i/\partial x = \partial n_e/\partial x=\partial \phi/\partial x=0$ at $x=0$. In order to maintain symmetry, we enforce $\phi=0$ at both $x=L$ and $x=-L$, even though we only solve for the positive half of the domain. This is accomplished through the output transform as described in Sec. \ref{sec:PIML} above.

A well established challenge when applying numerical ODE integration to Eqs. (\ref{eq:218}) and (\ref{eq:219}) is that a singularity exists for a finite ion temperature, where the solution is discontinuous about $(u_i/C_s)^2=T_i/T_e$. Strategies for overcoming this challenge involve including time dependence and converging to steady-state \cite{alvarez2020plasma} or writing density and velocity as unknown functions that depend on the bulk plasma conditions and Taylor expanding about the center of the plasma bulk \cite{crespo2018positive}. Another common singularity occurs where the velocities are set to zero, often in the center of the domain or where the plasma is set to be quasineutral. Not only does that result in a singularity when using ODEs, a traditional solver will find a trivial solution. To overcome this, traditional approaches start the simulation a small distance from that point and use backwards differencing to find the initial values at the new starting point ~\cite{riemann2005plasma,beving2022sheath, holmes2007introduction, valentini1988removal, sternberg1996approximation}. The PINN deals with these singularities by enforcing even parity on density and potential, which follows from the assumption of symmetry. Specifying boundary conditions at the far wall forces a non-trivial solution and negates the finite ion temperature singularity. 
\subsubsection{\label{sec:CS2}Results}

\begin{figure}
\centering
\subfigure[]{\includegraphics[scale=0.5]{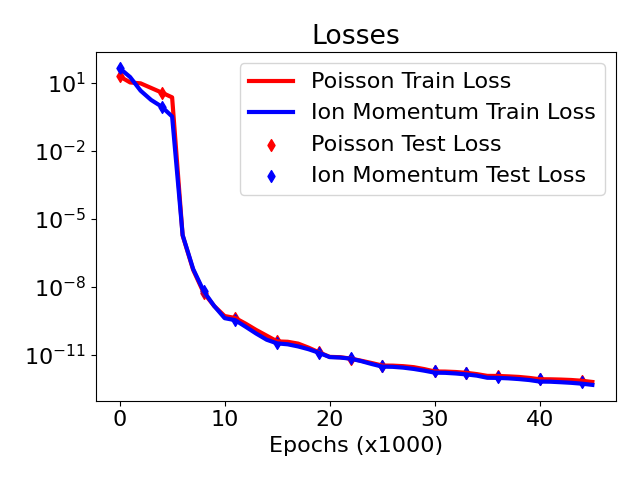}}
\subfigure[]{\includegraphics[scale=0.5]{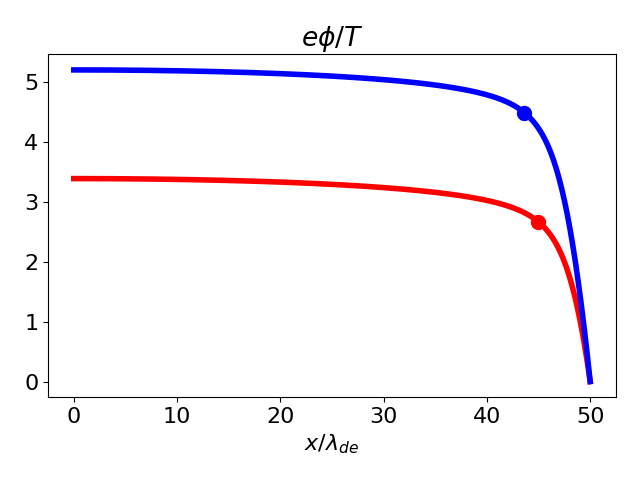}}
\subfigure[]{\includegraphics[scale=0.5]{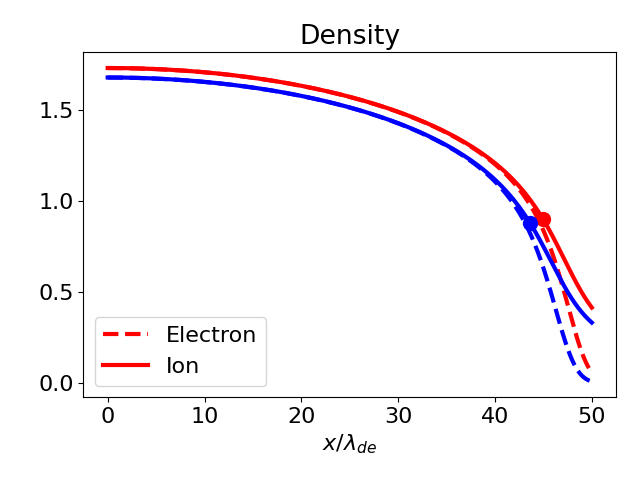}}
\subfigure[]{\includegraphics[scale=0.5]{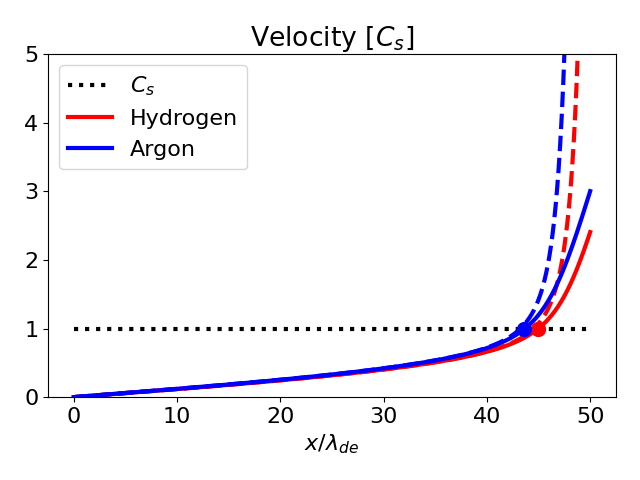}}
\caption{(a) The training and test loss, (b) electric potential, (c) densities, and (d) velocities. Two ions are demonstrated: Hydrogen (red) and Argon (blue). Ion temperature is zero for these figures and the plasma is assumed collisionless. The black dotted line represents the ion sound speed. The dots represent the sheath entrance.}
\label{fig:411}
\end{figure}

There are three free parameters that characterize the solution for the model described in Sec. \ref{sec:MCS}: the electron-to-ion mass ratio, $m_e/m_i$, the ion-to-electron temperature ratio, $T_i/T_e$, and the collisionality, $\lambda_{De}/\lambda_{in}$. The mass ratio is trained from hydrogen to argon, the temperature ratio is trained between 0 and 1, and the collisionality is trained for the range 0 to 0.3. With the addition of space, this becomes a 4 dimensional domain. The system size $L_x$ emerges as a fourth parameter, though for sufficiently large $L_x$, the sheath structure is largely insensitive to the system size, hence we will fix $L_x=50 \lambda_{De}$. When training the model, we used a fully connected feedforward neural network with 200,000 training points, a Hammersley distribution, four inputs, four hidden layers, each with 32 neurons, and two outputs, plus input and output transforms, to solve Eq. (\ref{eq:Model1}). The code is written using the DeepXDE library~\cite{lu2021deepxde}, with PyTorch as the backend. The ADAM optimizer is used for the first five thousand iterations and SSBroyden for the remainder. As shown in Fig. \ref{fig:411}, the loss for each equation reaches $5\times10^{-13}$, suggesting that the system is satisfied precisely, where an independent set of test points [solid markers in Fig. \ref{fig:411}(b)] are used to verify accuracy away from training points. 

As an independent means of verifying the solution, we have implemented a Runge-Kutta solver to directly solve Eq. (\ref{eq:Model1}). The detailed implementation of this ODE solver is described in Appendix \ref{sec:AppConstant}. As evident in Fig. \ref{fig:412}, the sheath solution found from a Runge-Kutta solver (denoted by `RK45' in the figure) is in excellent agreement with the prediction of the PINN. A small discrepancy is evident in the density profiles near $x=0$ in Fig. \ref{fig:412}(b), which arises due to the Runge-Kutta solver struggling with the singularity located at $x=0$.

\begin{figure}
\centering
\subfigure[]{\includegraphics[scale=0.5]{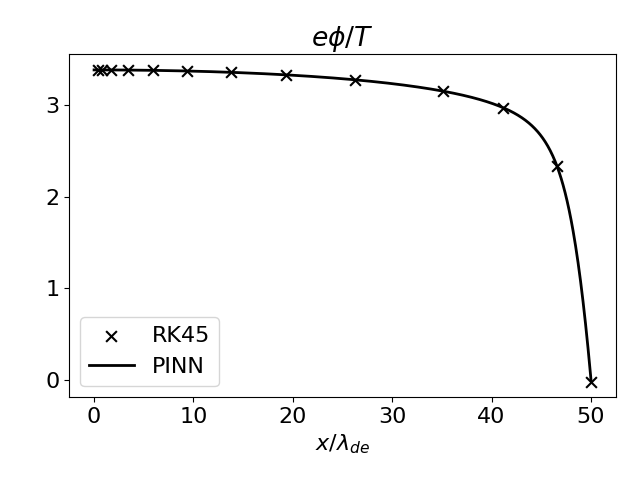}}
\subfigure[]{\includegraphics[scale=0.5]{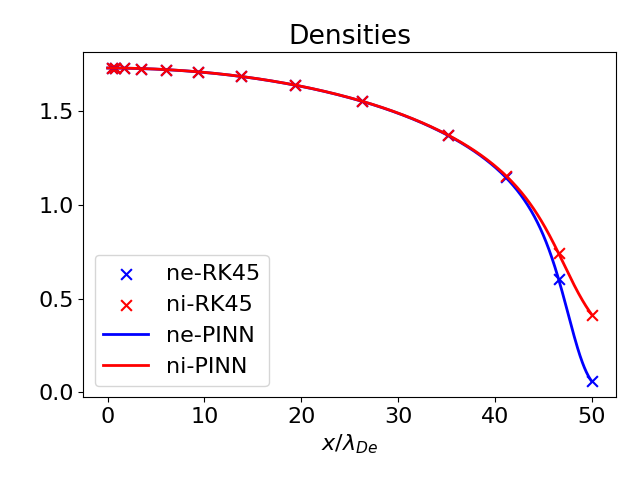}}
\subfigure[]{\includegraphics[scale=0.5]{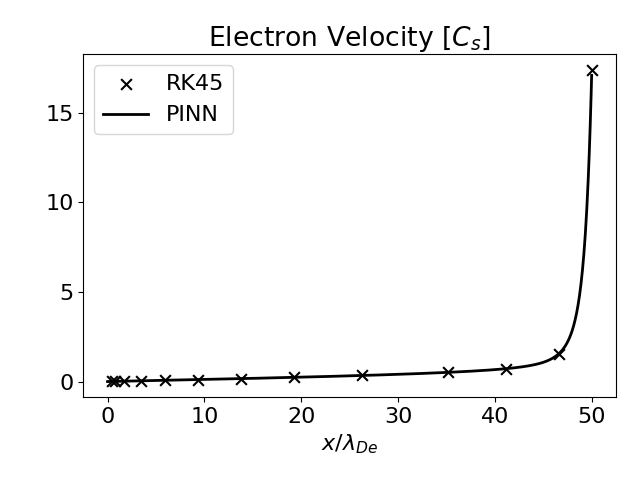}}
\subfigure[]{\includegraphics[scale=0.5]{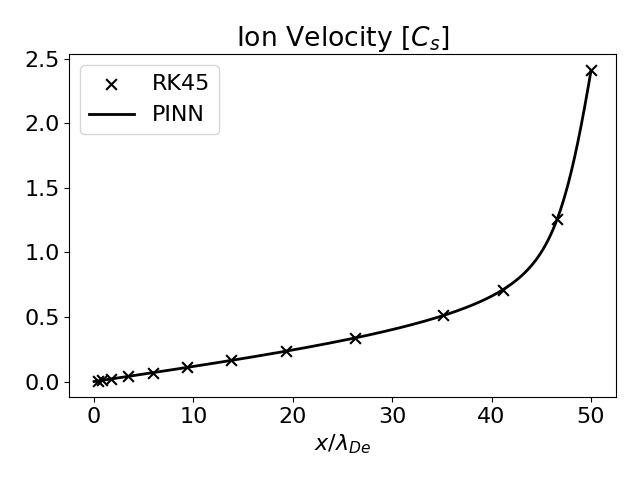}}
\caption{(a) The potential profile, (b) the density profiles for electrons and ions, (c) the electron velocity profiles, and (d) the ion velocity profile for collisionless hydrogen plasma with cold ions. Comparison between the RK45 prediction and the PINN prediction demonstrates that the PINN solved the equations accurately.}
\label{fig:412}
\end{figure}

\begin{figure}
\centering
\subfigure[]{\includegraphics[scale=0.5]{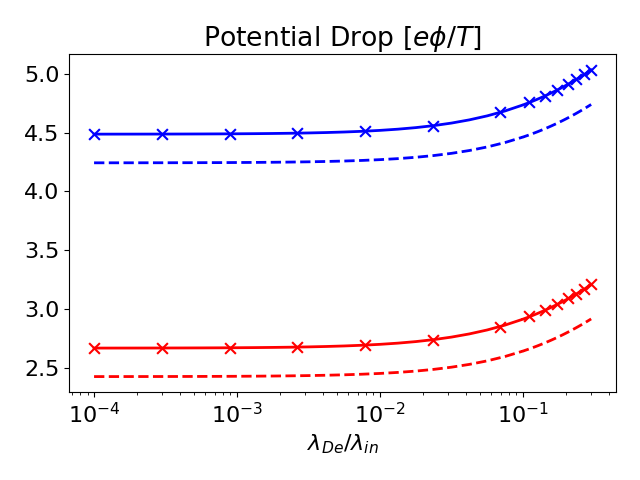}}
\subfigure[]{\includegraphics[scale=0.5]{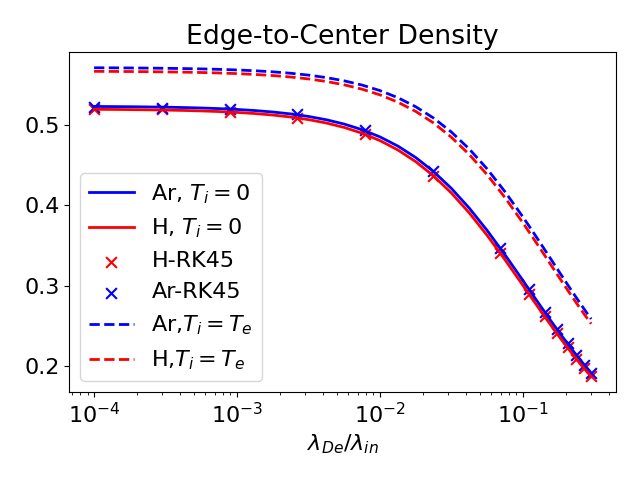}}
\subfigure[]{\includegraphics[scale=0.5]{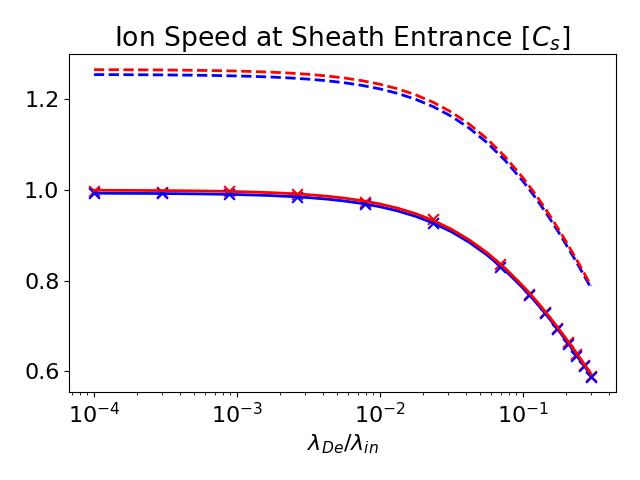}}
\subfigure[]{\includegraphics[scale=0.5]{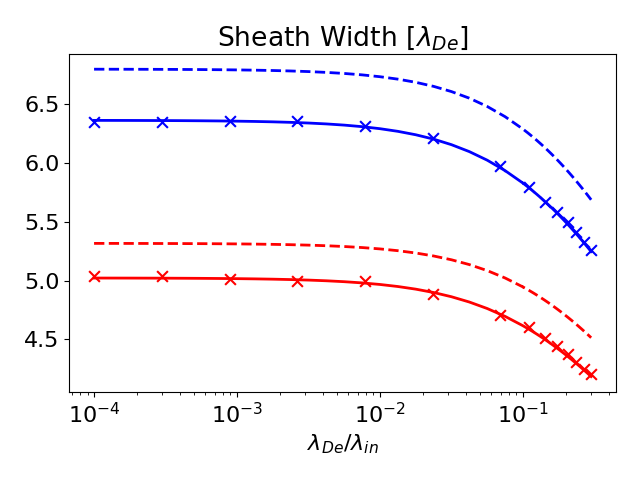}}
\caption{(a) The potential drop across the sheath. (b) The ratio of sheath entrance density to center density. (c) The ion speed at the sheath entrance. (d) The width of the sheath as a function of collisionality. The solid lines and the RK45 results are at $T_i=0$ and the PINN dotted lines are at $T_i=T_e$.}
\label{fig:413}
\end{figure}

Noting that the present PINN has been trained across a broad range of parameters, we can use the PINN to quickly identify how quantities such as the mass and temperature ratios, along with collisionality impact sheath properties. To do so, it will be useful to introduce an unambiguous definition of the sheath entrance. While the Bohm criterion is often used to define this point, we will be interested in understanding how the ion flow speed changes with system parameters, hence we will introduce a criterion distinct from the ion flow speed itself. An alternative definition that will allow for the ion flow speed dependence at the sheath entrance to be tracked is to use the fractional charge separation $\rho \equiv (n_i-n_e)/n_e$ as our definition of the entrance. To select a specific value of $\rho$, we choose the value when the ion flow equals the sound speed for a cold, collisionless hydrogen plasma. We find a value of $\rho = 0.0685$, which is indicated by the solid circle markers in Fig. \ref{fig:411}. With this definition, the dependence of several sheath properties on ion-neutral collisions is shown in Fig. \ref{fig:413}. Specifically, the potential drop across the sheath, the ratio of center density to sheath entrance density, the ion speed at the sheath entrance, and the width of the sheath. Figure \ref{fig:413} demonstrates the trends for the quantities for Hydrogen and Argon at $T_i=0$ and $T_i=T_e$ as a function of collisionality. These figures represent 200 predictions from the PINN and 50 from the RK45 method. We find that the predictions of the PINN and RK45 method agree precisely across the range of collisionalities. As the plasma becomes more collisional, the potential drop increases. Collisions act as a drag opposing the potential force. However, higher ion temperature increases the pressure at the center, which reduces the potential force necessary to oppose the drag force. Despite the slight reduction of the potential due to finite ion temperature, the ion species plays a larger role in determining the potential drop as the heavier species' inertia resists the potential and pressure forces. 

The density ratio, often referred to as the ``$h_l$ factor" in other literature \cite{raimbault2008edge,beving2022sheath, alvarez2020plasma}, is the ratio of density between the sheath edge and the center of the domain, $h_l=n_{edge}/n_{center}$. This ratio is often assumed in other models to be 0.5 \cite{robertson2013sheaths} in the low collisionality limit, which is a rounded value due to various definitions of the sheath edge, though \cite{beving2022sheath} uses PIC data to find a value of about 0.55 in the same limit. We find a value of 0.52, which agrees well with other fluid models, but not as well as a kinetic model, which is expected since the kinetic model includes additional physics. Rather than assume the density ratio, we allow the PINN to solve for the densities at the center and sheath entrance, thus self-consistently finding the $h_l$ factor for the entire parameter space. Figure \ref{fig:413}(b) demonstrates how increasing collisionality dramatically reduces the density ratio, while increasing ion temperature produces a modest increase. However, the ion species has minimal effect on the $h_l$ factor. We also find good agreement in the cold ion limit between the RK solver and the PINN.

At larger ion temperatures, the ion speed at the sheath entrance increases. As the plasma becomes more collisional, the ion speed decreases, regardless of the ion temperature. In the cold ion limit, we find good agreement with the RK solver across the range of collisionalities. Finally, as Figure \ref{fig:413}(d) demonstrates, the ion species plays a significant role in determining the width of the sheath. Heavier species require a larger potential, which causes the sheath to widen. Finite ion temperature further increases the sheath width as the fractional separation of ions and electrons occurs further from the wall due to the more energetic ions maintaining a greater density as they approach the wall as compared to the cold ion case. Conversely, the sheath width decreases with increasing collisionality, which is a consequence of the collisions inhibiting the ions from traveling to the wall. 

\begin{figure}
\centering
\subfigure[]{\includegraphics[scale=0.5]{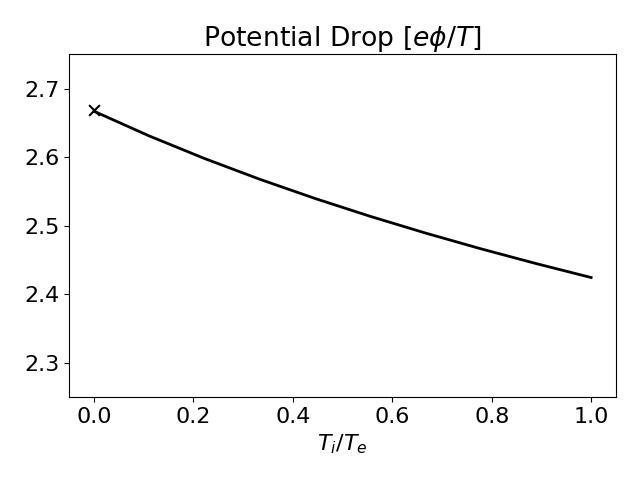}}
\subfigure[]{\includegraphics[scale=0.5]{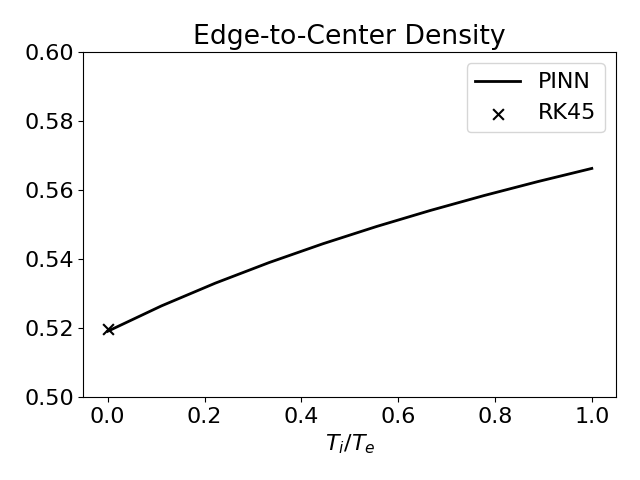}}
\subfigure[]{\includegraphics[scale=0.5]{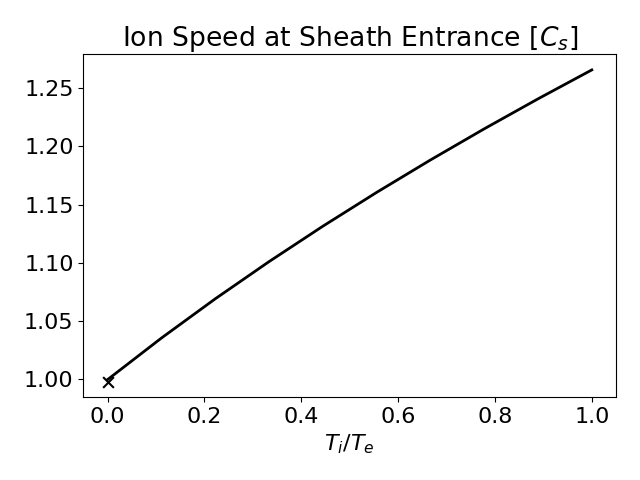}}
\subfigure[]{\includegraphics[scale=0.5]{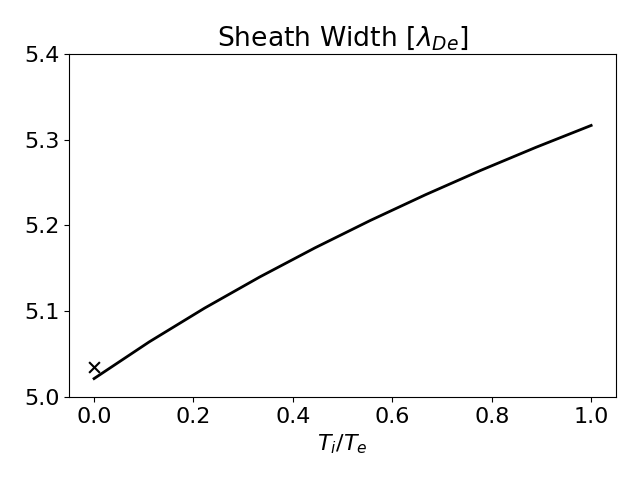}}
\caption{ (a) The potential drop across the sheath, (b) the ratio of sheath entrance density to center density, (c) the ion speed at the sheath entrance, and (d) the width of the sheath as a function of temperature ratio. The "x" indicates the RK45 solution when $T_i=0$, since it is singular for finite ion temperature. }
\label{fig:414}
\end{figure}

Because the numerical approach is singular for finite ion temperature, Figure \ref{fig:414} demonstrates how the ion temperature affects the integrated quantities according to the PINN and how the PINN recovers the RK45 solution as ion temperature goes to zero. The ``x" indicates the RK45 solution when $T_i=0$ and the black curve demonstrates how the ion temperature affects the integrated quantities. Figure \ref{fig:414}(a) indicates how the potential drop decreases as the energetic ions have a larger pressure force pushing them towards the wall, thus a weaker potential force is needed. Increasing ion temperature reduces the center density as the increased pressure pushes ions towards the wall. As a result, the edge-to-center density increases with ion temperature [see Fig. \ref{fig:414}(b)]. The ion speed at the sheath entrance increases with increasing ion temperature, as indicated in Fig. \ref{fig:414}(c) There is a corresponding widening of the sheath due to this effect, which Figure \ref{fig:414}(d) demonstrates.

\subsection{\label{sec:Complex}Self-Consistent Ionization Source}

\subsubsection{\label{sec:Modelcomplex2}Model}

A more complete description of the plasma-sheath transition region can be achieved by including self-consistent collisional ionization and recombination sources. The two kinds of recombination considered are radiative recombination and three-body recombination. There are three main factors that determine the impact of collisional ionization and recombination on the sheath profile: electron temperature, density, and domain size. For a very small domain, collisional ionization and recombination will often have minimal impact on the sheath solution. Additionally, for low temperatures in the range 1 eV to 20 eV, collisional ionization and both types of recombination considered are negligible compared to the constant source rate, whereas for the higher temperatures, collisional ionization becomes comparable to the constant source rate, while both radiative recombination and three-body recombination remain negligible. Thus, we will neglect both kinds of recombination, though we will include collisional ionization. The inclusion of this term will substantially complicate the solution of the model equations, since the source term appearing in the continuity equations will contain explicit dependence on the plasma density and electron temperature. We will no longer be able to analytically integrate the continuity equations, as was done in Sec. \ref{sec:CS} above, and will need to solve a substantially more complex set of equations. Considering the collisional ionization source will thus be useful to demonstrate that the PINN approach adopted in this paper is able to treat this more complex system. 

We again consider a symmetric plasma that contains a neutral background of argon and two charged species: electrons and singly ionized argon, that are treated as fluids. These species are considered at constant temperature, though the magnitude of the electron and ion temperatures are inputs into the model, where we will learn the solution for a broad range of temperatures. The computational domain is the same as used in Sec. \ref{sec:CS}, extending from $x=0$ to $x=L_x=50 \lambda_{De}$. Treating the electrons adiabatically as defined in Equation \ref{eq:217}, the equations for this model are: 
\begin{subequations}
\label{eq:ComplexSource}
\begin{equation}
    \label{eq:221} 
    \epsilon_0\frac{\partial^2\phi}{\partial x^2} = -e(n_i-n_e),
\end{equation}
\begin{equation}
    \label{eq:222} 
    \frac{\partial}{\partial x}(n_iu_i) = S_0 + S_{ion},
\end{equation}
\begin{equation}
    \label{eq:223} 
    \frac{\partial}{\partial x}(n_eu_e) = S_0 + S_{ion},
\end{equation}
\begin{equation}
    \label{eq:225} 
    m_in_iu_i \frac{\partial u_i}{\partial x} + T_i\frac{\partial n_i}{\partial x} = -en_i\frac{\partial \phi}{\partial x} - m_iu_i(S_0 + S_{ion}) + R_{in},
\end{equation}
\end{subequations}
with the collisional ionization rate described by $S_{ion} = n_nn_eS_{ionize}$, and

\begin{equation}
\label{eq:226}
    S_{ionize} = 10^{-11}\frac{(T_e/E_z)^{(1/2)}}{(E_z)^{(3/2)}(6.0+T_e/E_z)}e^{\left(\frac{-E_z}{T_e}\right)}\;\left[ \text{m}^3/\text{sec}\right].
\nonumber
\end{equation}
Here, $T_e$ is the electron temperature in eV, $E_z$ is the ionization potential in eV, and $S_{ionize}$ describes collisional ionization rate coefficient. The collisional ionization model that we will employ represents fits to cross section data and is taken from the NRL Formulary ~\cite{richardson20192019}. We note that more accurate ionization rates could easily be used, though we will find this form to be convenient for this work.  

While the introduction of collisional ionization will prevent the continuity equations from being integrated analytically, the electron momentum equation can still be integrated to give Boltzmann electrons. We will also make use of the ambipolarity condition $n_iu_i = n_eu_e$ to eliminate one of the continuity equations, which allows us to use the algebraic relation $u_i = n_eu_e/n_i$ to constrain the ion velocity. Once again writing in residual form and normalizing, the system becomes:
\begin{subequations}
\label{eq:model2}
\begin{equation}
    \label{eq:228} 
    0 = (n_e-n_i) - \frac{\partial^2\phi}{\partial x^2},
\end{equation}
\begin{equation}
    \label{eq:229} 
    0 = \frac{1}{L} + n_nn_e\bar{S}_{ionize}\bar{S}_0-\frac{\partial}{\partial x}(n_eu_e) ,
\end{equation}
\begin{equation}
    \label{eq:2210} 
    0 = -n_i\frac{\partial \phi}{\partial x} - u_i\left(\frac{1}{L} + n_nn_e\bar{S}_{ionize}\bar{S_0} \right) - n_nn_iu_i\sigma_{in} -n_iu_i \frac{\partial u_i}{\partial x} - \frac{T_i}{T_e}\frac{\partial n_i}{\partial x},
\end{equation}
\end{subequations}
with the electron density taken to be Boltzmann as defined by Eq. (\ref{eq:217}) above, the neutral density as an input parameter, $n_n$, the normalized ionization rate coefficient, $\bar{S}_{ionize}=S_{ionize}/(\lambda_{De}^2C_s)$, and the normalized source rate, $\bar{S_0}=S_0\lambda_{De}^4L/C_s$. The singly ionized Argon to neutral Argon collisional cross-section is taken to be isotropic, is evaluated at the ion sound speed, and is described by:
\begin{equation}
    \sigma_{in} = \left(\frac{2\times 10^{-19}}{\sqrt{2E}(1+2E)} + \frac{6\times 10^{-19}E}{(1+\frac{2}{3}E)^{2.3}}\right)\lambda_{De}n_0,
\end{equation}
where $E$ is the energy, up to 10 keV, corresponding to the sound speed, $E=0.5T_e(1+T_i/T_e)$. This fit is taken from the Phelps database provided by LXCat \cite{Phelps}. Compared to the previous model, the ion neutral collisionality is no longer a parameter that is scanned over: the inputs controlling the collisionality are the neutral density and the electron temperature. 

Additional complexity is introduced because the continuity equation [Eq. (\ref{eq:216})] can no longer be integrated due to the collisional ionization source such that the value of $n_w$ is not known. We allow the network to solve for $n_w$ as a trainable variable, thus adding $n_w$ to the other outputs: $\phi$, $n_i$, and $u_e$. The impact of collisional ionization depends sensitively on the system size as evident from Eq. (\ref{eq:2210}). For larger systems, the constant source term, which in normalized units has the form $1/L$, decreases in magnitude relative to collisional ionization. In addition, and somewhat counterintuitively, as the strength of the constant source $S_0$ is increased, collisional ionization becomes increasingly important. The reason for this latter trend is that the ionization scales with the square of density, such that for a given neutral fraction $n_n/n_0$ a stronger particle source leads to a nonlinear increase in the magnitude of the ionization term. 

The introduction of the ionization term will also require the electron temperature to be explicitly specified. We will scan over a range of electron temperatures, as well as ion to electron temperature ratios. This model also introduces another free parameter, $\bar{S}_0$, associated with the constant source strength. Finally, the last free parameter is the neutral density, which will influence both the ionization strength and ion collisional drag.

The boundary conditions for this model are similar to the previous model. The assumption of symmetry allows: $u_i=u_e=0$ and $\partial n_i/\partial x = \partial n_e/\partial x=\partial \phi/\partial x=0$ at $x=0$.
In order to preserve symmetry, we enforce $\phi = 0$ at both walls, allowing the potential to float in the center of the domain. Finally, we require that the electron velocity at the walls to be (in normalized units reads):
\begin{equation}
    u_{e,wall} = \sqrt{\frac{m_iT_e}{2\pi m_e(T_e+T_i)}},
\end{equation}
at $x=L$. Conversely, at $x=-L$, the electron velocity is enforced as $-u_{e,wall}$ to preserve proper parity. 

\subsubsection{\label{sec:complex2}Results}
 
The model is five dimensional with one spatial dimension and four physical parameters. We used the PyTorch library \cite{paszke2019pytorch} for this model. The electron temperature is varied between 1 eV and 20 eV, $T_i/T_e$ is varied between 0 and 1, $S_0$ is varied between $10^{28}$ and $10^{29} \text{ m}^{-3}\text{s}^{-1}$, and finally neutral density is varied between 0 and 5. We find it convenient to use the dimensional source rate as an input since the normalizations we chose are allowed to vary with the input source, rather than choosing an arbitrary reference density. 

For this case, we found that SSBroyden provided a robust optimization algorithm for training. The network consisted of 5 hidden layers, each with 32 neurons and two million training points distributed according to a Sobol distribution were used for the domain. Additionally, one hundred thousand points were distributed on each boundary. Increasing the network size and number of points, as compared to the previous model, directly results from the increased complexity of this model. The ionization source increases the non-linearity of the problem and requires the network to be more expressive in order to converge.
For simplicity, the PINN was trained for singly ionized argon, though more ion species could be included if needed for a specific application, provided the appropriate cross-section data is also included. However, adding more species requires an additional set of equations per species in the loss function and additional collision terms between species, which adds complexity to the model. If the added complexity requires a larger network or more points to converge, the training time is expected to increase. One strategy for dealing with large numbers of loss terms is adaptive weighting, which automatically balances the loss terms during training so the network converges them equally \cite{wang2021understanding}, which can also reduce the network size and number of points needed to converge, thereby speeding up the training. For this model, using the procedure described in Sec. \ref{sec:CS2} above, we found $\rho = 0.0685$ sets the fractional charge separation at the location that we will define as the sheath entrance.

\begin{figure}
\centering
\includegraphics[scale=0.5]{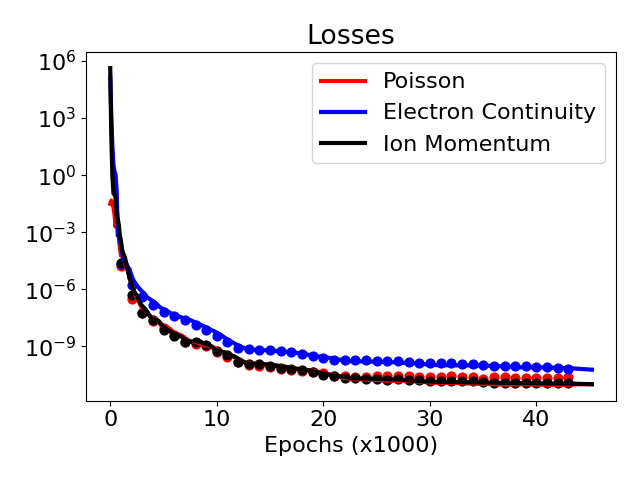}
\caption{Loss history for the sheath model defined by Eq. (\ref{eq:model2}). The solid lines are the training loss and the points are the test loss.}
\label{fig:421sub0}
\end{figure}

\begin{figure}
\centering
\subfigure[]{\includegraphics[scale=0.5]{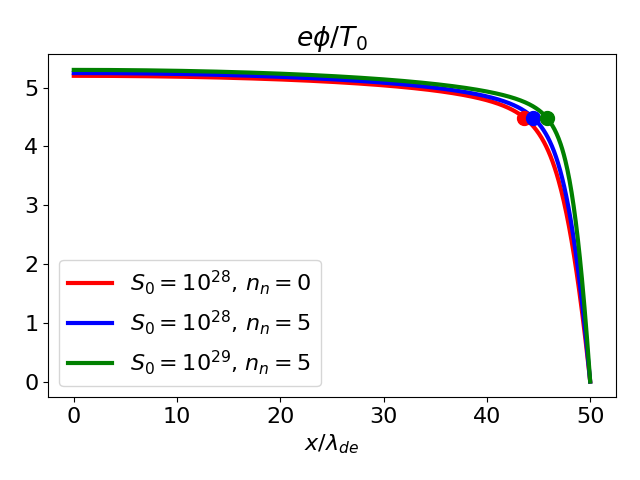}}
\subfigure[]{\includegraphics[scale=0.5]{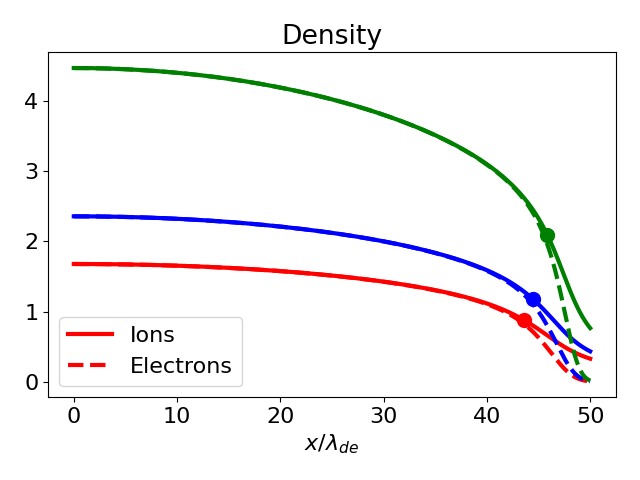}}
\subfigure[]{\includegraphics[scale=0.5]{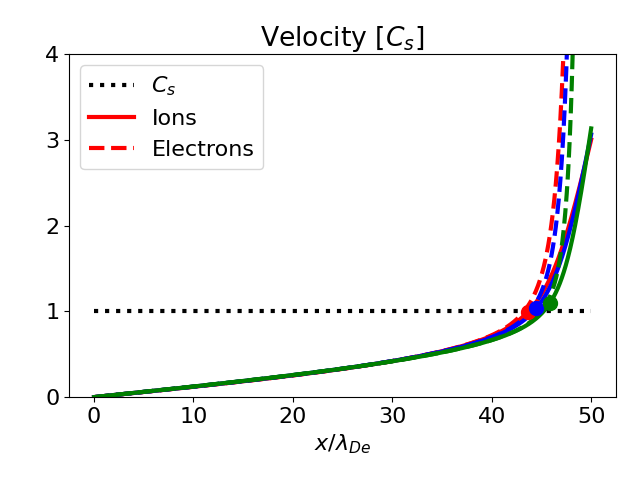}}
\subfigure[]{\includegraphics[scale=0.5]{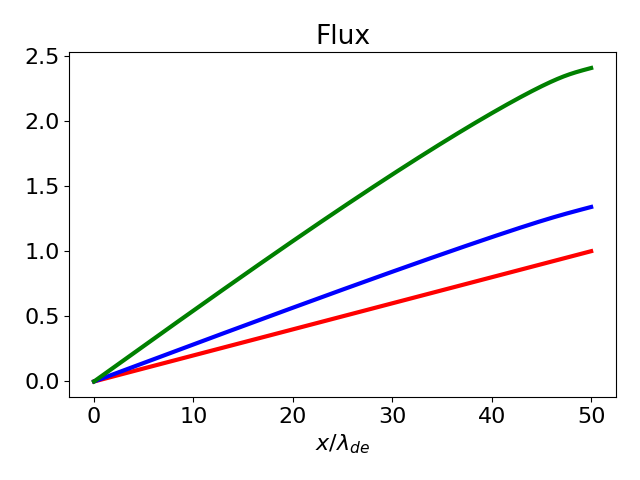}}
\caption{(a) The potential profile, (b) the density profiles for ions and electrons, (c) the velocity profiles for ions and electrons, and (d) the flux profile for various source strengths and neutral densities. In each case, $T_e=20\;\text{eV}$ and $T_i=0$. The green lines indicate the strongest ionization source, whereas the red lines indicate no ionization. The source rate, $S_0$, is in units of $\text{ m}^{-3}\text{ s}^{-1}$ and the neutral density, $n_n$, is normalized. The dots indicate the locations of the sheath entrance according to the value of $\rho=0.0685$.}

\label{fig:421}
\end{figure}

Figures \ref{fig:421sub0} and \ref{fig:421} show the loss history, example potential, velocity, density, and flux profiles across the domain for a variety of plasma parameters. While beginning at a higher initial loss compared to the simpler model discussed in Sec. \ref{sec:CS2} above, the loss history drops by about fifteen orders of magnitude. The final loss reaches $ 10^{-10}$, indicating the PINN has been able to find a solution that satisfies Eq. (\ref{eq:model2}) to a high degree of accuracy. For the small system size considered here, fifty Debye lengths, the potential and velocity profiles are weakly impacted by the ionization source. However, increasing the source strength and neutral density leads to a significant change in the total density. In the collisionless case without neutrals, the center density is only $3 \times 10^{20}\text{ m}^{-3}$, while the center density is $3.7 \times 10^{21}\text{ m}^{-3}$ in the highest source case. This is due to the normalizations changing with the input source strength. In the strongest ionization source case, the particle flux is more than two times the flux in the collisionless case, implying that the ionization source dominates over the constant source. This indicates that the PINN solves for a parameter space where the ionization source is non-trivial. 

\begin{figure}
\centering
\subfigure[]{\includegraphics[scale=0.5]{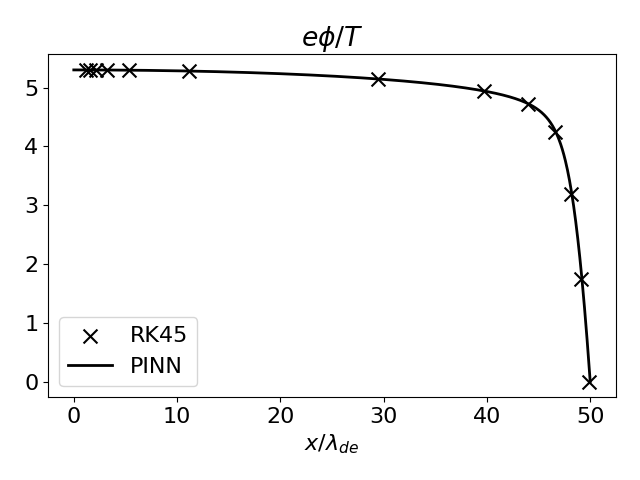}}
\subfigure[]{\includegraphics[scale=0.5]{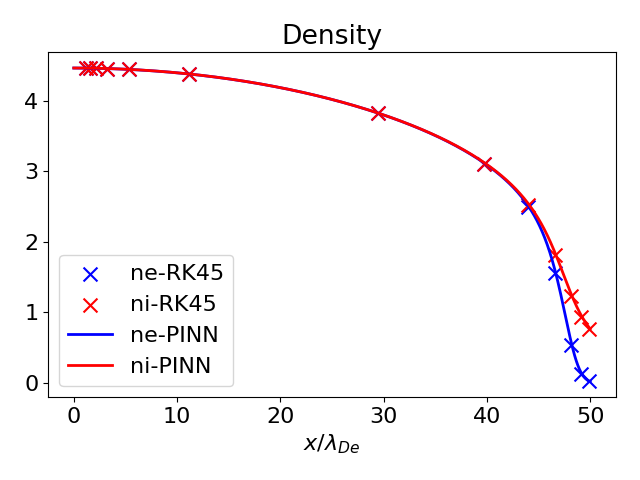}}
\subfigure[]{\includegraphics[scale=0.5]{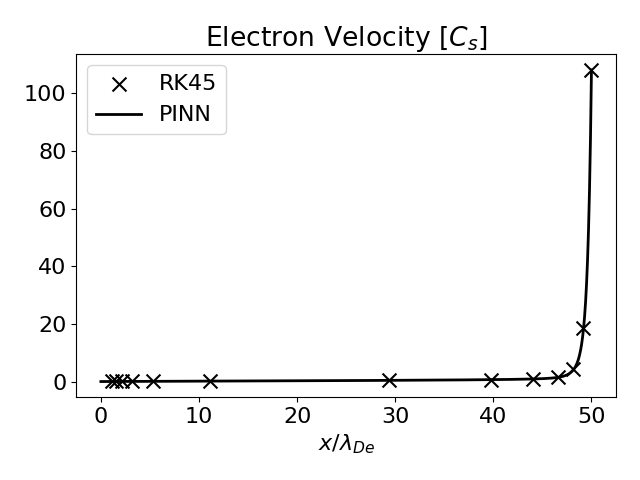}}
\subfigure[]{\includegraphics[scale=0.5]{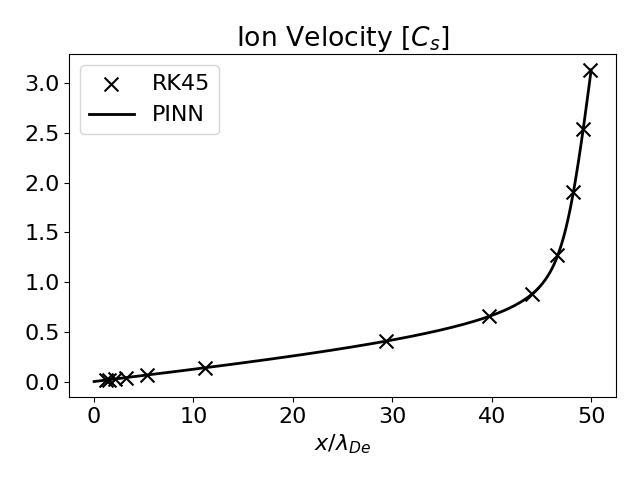}}
\caption{(a) The potential profile, (b) the density profiles for ions and electrons, (c) the electron velocity profile, and (d) the ion velocity profile comparing the PINN and RK45 methods for argon with $S_0=10^{29}\text{ m}^{-3}\text{s}^{-1}$, $T_e=20\text{ eV}$, the normalized neutral density: $n_n=5$, and cold ions. The RK45 predictions are denoted by "x" and the solid lines are the PINN predictions.}
\label{fig:422}
\end{figure}

To validate our PINN solution, we have again implemented a Runge-Kutta solver, where Fig. \ref{fig:422} demonstrates that the predictions from the PINN and the Runge-Kutta solver are again in excellent agreement. The case selected is the profile with the largest ionization rate included in the range of training, $T_e=20 \text{ eV}$, $T_i=0$, $S_0=10^{29}\text{ m}^{-3}\text{ s}^{-1}$, and $n_n=5$. In this case, due to the large gradient in the electron velocity caused by the assumption of the electron velocity at the wall, a Radau solver was used, rather than the RK45 method used in the previous model. Radau is a fully implicit Runge-Kutta solver that is more stable and efficient for stiff problems. Figure \ref{fig:422} demonstrates excellent agreement between the PINN and the Radau solver for all profiles. 

\subsection{\label{sec:HE}Heat Equation}

\subsubsection{\label{sec:modelHE2}Model}

The last fluid model we will consider incorporates an electron heat equation, and thus allows us to relax the constant temperature assumption made in the previous models. Relaxing the constant electron temperature assumption will prevent the use of Boltzmann electrons, such that both the electron momentum and heat equations must be added to the model. For this section, we will begin by assuming cold ions, eliminating the need for an ion heat equation. We will also assume a constant source of particles and heat, and, for simplicity, neglect ion-neutral collisions.

To obtain a finite steady state electron temperature, we will include a constant heating source of the form $S_{Te}=3T_sS_0/2$, where $T_s$ is the source temperature, and $S_0$ is the magnitude of the particle source. Noting that the electron temperature will now vary across the domain, we will use the source temperature $T_s$ as the temperature normalization, yielding the system of normalized equations

\begin{subequations}
\label{eq:heatequations}
\begin{equation}
    \label{eq:231} 
    0 = (n_e-n_i)-\frac{\partial^2\phi}{\partial x^2},
\end{equation}
\begin{equation}
    \label{eq:232} 
    0 = \frac{m_i}{m_e}n_e\frac{\partial \phi}{\partial x} - \frac{u_e}{L}+R-\frac{m_i}{m_e}\frac{\partial}{\partial x} \left( n_eT_e \right)
    ,
\end{equation}
\begin{equation}
    \label{eq:234} 
    0 = -n_i\frac{\partial \phi}{\partial x} - \frac{u_i}{L} -R-\frac{x}{L} \frac{\partial u_i}{\partial x},
\end{equation}
\begin{equation}
    \label{eq:235} 
    0 = \frac{1}{2L}\left(3-{3T_e+\frac{m_e}{m_i}u_e^2}\right) +Q-\frac{3}{2}\frac{x}{L}\frac{\partial T_e}{\partial x} - n_eT_e\frac{\partial u_e}{\partial x} - \frac{\partial q}{\partial x},
\end{equation}
\end{subequations}
where we have integrated the ion and electron continuity equations yielding $n_eu_e=n_iu_i=x/L$, the momentum exchange between ions and electrons is $R = R_{u} + R_{T}$, $q$ is the conductive heat flux, which is given by $q=0.71n_eT_e(u_e-u_i)-\kappa_0\frac{\partial T_e}{\partial x}$, where the heat conductivity is given by $\kappa_0 = 3.16n_eT_e\tau_{ei}/m_e$ with $\tau_{ei}$ defined below~\cite{braginskii1965transport}, and $Q=Q_R+Q_{eq}$. The first term in Eq. (\ref{eq:235}) is the plasma heating term due to the constant source. If the sum of the electron thermal energy and electron kinetic energy is less than the source, then the source heats the plasma, and vice versa. $R_{u}$ is the momentum change due to electrons colliding with ions, and $R_{T}$ is the thermal force. Their definitions are given by:
\begin{subequations}
\begin{equation}
\label{eq:236}
    R_{u} = \frac{-m_en_e}{m_j\tau_{ei}}0.51(u_e-u_i),
\end{equation}
\begin{equation}
    \label{eq:237}
    R_{T} = -\alpha n_e\frac{m_i}{m_j}\frac{\partial T_e}{\partial x},
\end{equation}
\begin{equation}
    \label{eq:238}
    \tau_{ei} = \frac{12\pi^{3/2}\sqrt{m_e/m_i}}{\sqrt{2} \Lambda} \frac{T_e^{3/2}}{Z^2n_i}\frac{\epsilon_0^2T_s^2}{q^4\lambda_{De}n_0},
\end{equation}
\end{subequations}
where $\alpha=0.71$ for hydrogen, but is generally an ion dependent parameter \cite{braginskii1965transport}, the subscript $j$ indicates ions or electrons and results from the normalization of the momentum equations using the electron and ion masses for the electron and ion momentum equations, respectively, and $\Lambda$ is the Coulomb logarithm. Next, the energy exchange between ions and electrons $Q_R$ and $Q_{eq}$, can be written:
\begin{subequations}
\begin{equation}
\label{eq:239}
        Q_R = -R(u_e-u_i),
\end{equation}
\begin{equation}
    \label{eq:2310} 
    Q_{eq} = \frac{-3m_en_e}{m_i\tau_{ei}}(T_e-T_i).
\end{equation}
\end{subequations}
In this formalism, $u_i = x/(Ln_i)$ and $u_e=x/(Ln_e)$. Thus, the network solves for $\phi, n_i, n_e,$ and $T_e$. Boundary conditions are: $\partial n_i/\partial x = \partial n_e/\partial x=\partial \phi/\partial x=\partial T_e/\partial x=0$ at $x=0$ and $\phi = 0$ and $n_e = \sqrt{2\pi m_e/(T_{e,wall}m_i)}$ at $x=L$. A trainable variable is used for the electron temperature, $T_e=T_{e,wall}$, at $x=L$. Since the heat equation is second order, another boundary condition is required on the temperature. Taking the steady state energy equations and assuming no ion energy source \cite{helander2005collisional}:
\begin{subequations}
\label{eq:energyequations}
\begin{equation}
    \label{eq:2311}
    \frac{\partial Q_e}{\partial x} = -en_eEu_e + \int d^3v\frac{1}{2}m_ev_e^2C_e + S_{T_e},
\end{equation}
\begin{equation}
    \label{eq:2312}
    \frac{\partial Q_i}{\partial x} = Zen_iEu_i + \int d^3v\frac{1}{2}m_iv_i^2C_i,
\end{equation}
\end{subequations}
where $Q_s$ is the flux of energy of species $s$ (not to be confused with the energy exchange term $Q$ defined above) and $S_{T_e}$ is a constant energy source. Due to the ambipolar field, no net current density flows to the wall, thus we have for singly charged ions, $j=(en_iu_i+en_eu_e)=0$, and the collisional energy exchange terms cancel out, i.e. energy lost by electrons due to collisions with ions goes to the ions and vise-versa. Summing Eqs. (\ref{eq:2311}) and (\ref{eq:2312}) results in:
\begin{equation}
    \label{eq:2315}
    \frac{\partial}{\partial x} (Q_e + Q_i) = S_{T_e}.
\end{equation}
Equation (\ref{eq:2315}) can be integrated for a constant source. Integrating over the range $x=0$ to $x=L_x$, $S_{Te}$ becomes $3T_sS_0L_x/2$, which, when normalized, becomes $3/2$. Using the definitions of $Q_e$ and $Q_i$ \cite{helander2005collisional},

\begin{subequations}
\begin{equation}
    \label{eq:2313} 
    Q_e = 0.71n_eT_e(u_e-u_i) - \kappa_0\frac{\partial T_e}{\partial x}+\left(\frac{5}{2}n_eT_e+\frac{1}{2}m_en_eu_e^2\right)u_e, 
\end{equation}
\begin{equation}
    \label{eq:2314}
    Q_i = \frac{1}{2}m_in_iu_i^3,
\end{equation}
\end{subequations}
where we have assumed cold ions for the ion energy equation, rearranging for $\partial T_e/\partial x$, and normalizing, we can write:

\begin{equation}
    \label{eq:2316} 
    \frac{\partial T_e}{
    \partial x}=\frac{\frac{3}{2}-\frac{5}{2}T_e-\frac{m_e}{2m_i}u_e^2 -\frac{1}{2}u_i^2 - 0.71n_eT_e(u_e-u_i)}{-3.16n_e\frac{m_i}{m_e}T_e\tau_{ei}}.
\end{equation}
Looking at the first two terms in the numerator, we can see that if $T_e$ is above 0.6, then the gradient is guaranteed to be positive, which is non-physical, as we expect $u_e \geq u_i$ throughout the domain. In particular, for a uniformly distributed heat source in the symmetric geometry shown in Fig. \ref{fig:21}, the heat flux should be conducted to the wall implying a negative electron temperature gradient. Thus, 0.6 was chosen as the ceiling for $T_{e}$ and $T_{e,wall}$ when training the PINN. Additionally, density is constrained to be positive definite. 

\subsubsection{\label{sec:HE2}Results}

\begin{figure}
\centering
\subfigure[]{\includegraphics[scale=0.5]{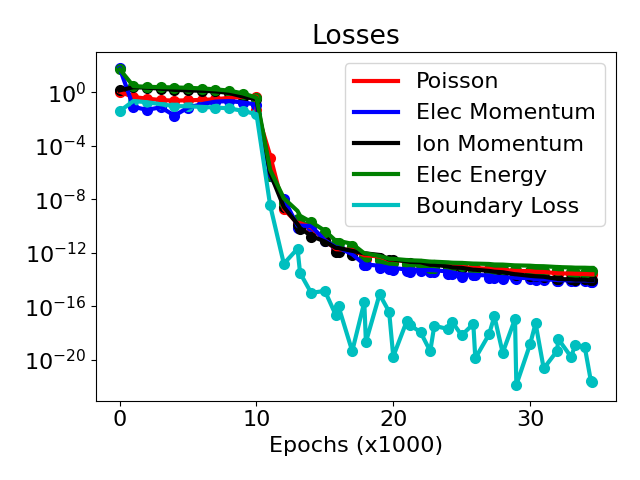}}
\subfigure[]{\includegraphics[scale=0.5]{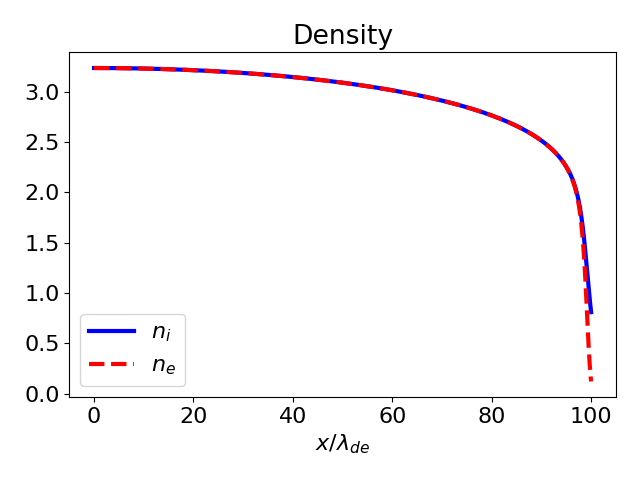}}
\caption{(a) The loss history and (b) the ion and electron density profiles of the of the model. The solid lines are training losses and the dots are test losses.}
\label{fig:431sub0}
\end{figure}

For this model, hydrogen is the ion species considered, with the model trained with ADAM for the first ten thousand iterations and SSBroyden for the remainder with one hundred thousand Hammersley distributed training points. The neural network contained three hidden layers of twenty neurons each in a fully connected feed-forward neural network. For simplicity, we did not train this model over the physical parameters of the model, such that the only input into the neural network was the independent variable $x$. This model was also trained for a high density ($n_0=8.3\times 10^{21}$ $m^3$). Considering the high collisionality limit, where the electron-ion mean free path is about 3 Debye lengths, allowed us to find a nontrivial electron temperature profile, whereas in the low-collisionality limit the electron temperature asymptotes to a constant value across the domain. Figure \ref{fig:431sub0} shows the training history of the model system given by Eq. (\ref{eq:heatequations}). The PINN is again able to achieve a very low loss, reaching $10^{-13}$ for the ODEs, and an even smaller loss for the boundary condition defined by Eq. (\ref{eq:2316}). Additionally, we chose a larger domain of $L_x=100\lambda_{De}$ to better illustrate the variation of the electron temperature, where the PINN is able to resolve the now very fine sheath forming near the edge of the domain.

\begin{figure}
\centering
\subfigure[]{\includegraphics[scale=0.5]{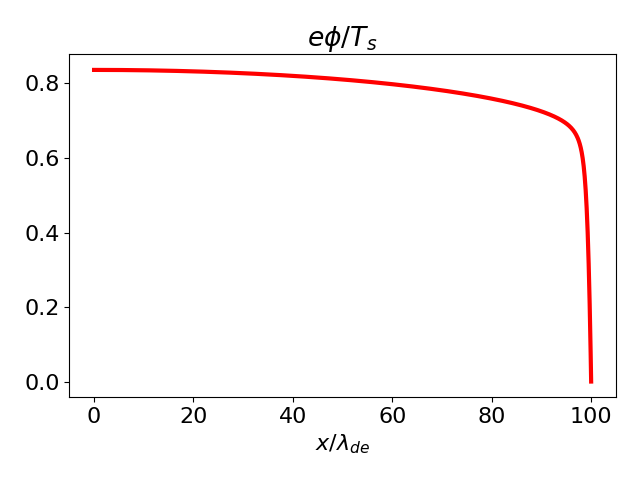}}
\subfigure[]{\includegraphics[scale=0.5]{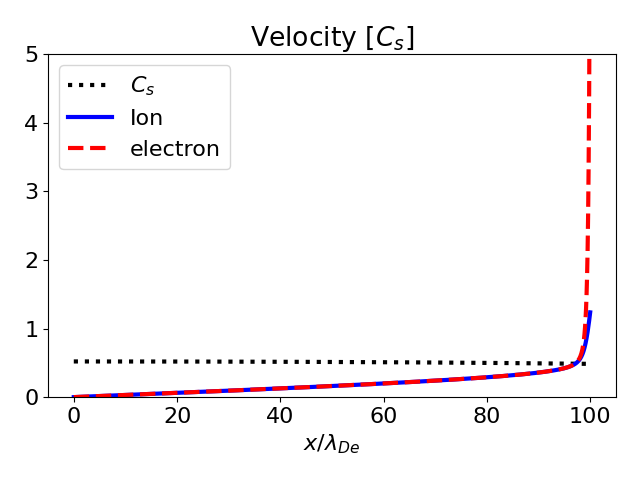}}
\subfigure[]{\includegraphics[scale=0.5]{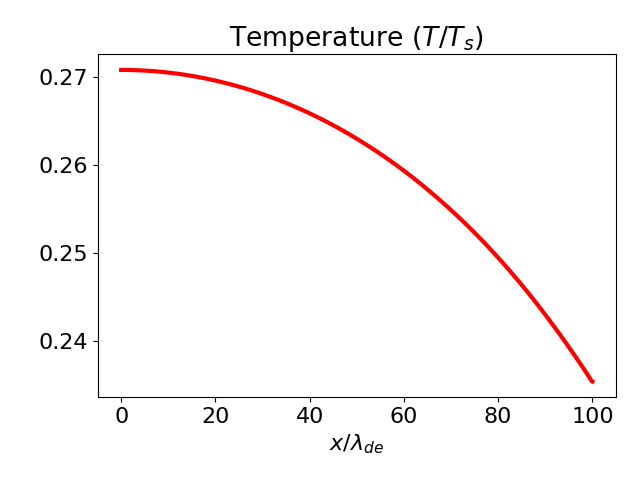}}
\subfigure[]{\includegraphics[scale=0.5]{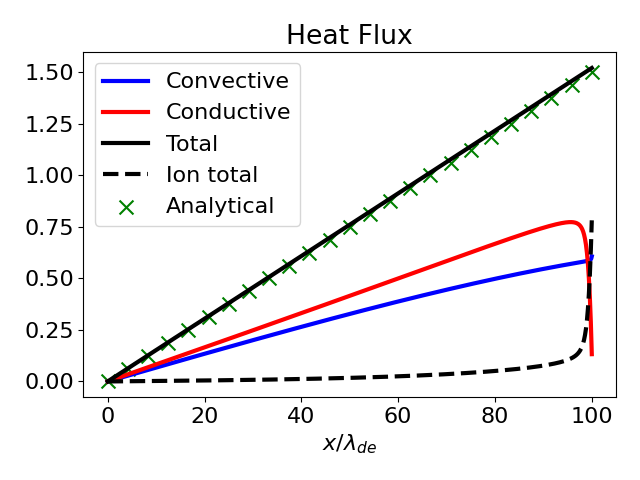}}
\caption{(a) The potential profile, (b) the ion and electron velocity profiles with the local (temperature dependent) sound speed, (c) the electron temperature profile, and (d) the heat flux profiles for a hydrogen plasma with cold ions. All quantities are normalized. The source temperature is $T_s=3$ eV and the source rate is $10^{31} \text{ m}^{-3}\text{ s}^{-1}$}
\label{fig:431}
\end{figure}

The plasma is quasineutral in the center and the densities separate in the sheath as a potential is formed. The apparent lower potential drop across the sheath $\approx 0.85T_s$ is due to the electron temperature being substantially lower than the source temperature $T_s$. Normalizing to the electron temperature at the center of the domain, the potential drop is 2.52 eV. The electron temperature decreases towards the wall, though only modestly even for the fairly large domain considered in this example. With such a low temperature and high density, the conductive losses are comparable to convective losses, but at higher temperatures or lower densities, conduction dominates. Specifically, the heat conductivity scales with $T_e^{5/2}$ and convection scales with $n_e$, such that we expect the largest temperature variation to emerge in the high density, low temperature regime.

At the sheath entrance, the electron distribution is composed broadly of two parts that can contribute to the heat transfer: a majority of forward-going slower electrons that will be repelled by the sheath, and a faster tail that will make it to the wall and maintain the sheath potential. This potential is how electrons transfer energy to the ions. The electron heat transmission coefficient can be calculated from the information presented in Figure \ref{fig:431}: $\gamma_e=q_{se}^e/(\Gamma_eT_e)$, where $q_{se}^e$ is the electron heat flux, $\Gamma_e$ is the electron particle flux, and $T_e$ is the electron temperature, and each quantity is evaluated at the sheath entrance, which is defined the same way as the previous models. The heat transmission coefficient is a measure of how much heat is lost in the sheath, including particles that exhaust on the wall, and was calculated as 5.8 for electrons, which is in reasonable agreement with the simple estimate of 5.5 from Ref. \cite{stangeby}. Figure \ref{fig:431} demonstrates how the electron heat flux decreases in the sheath exactly proportional to the increase in ion heat flux demonstrating how electrons are transferring energy to ions. The solid black line is the sum of the ion conductive heat transport (convection is negligible) and the electron conductive and convective heat transport and aligns well with the expected profile of the heat flux, marked with green exes. The analytical solution, $q=3x/(2L)$, is obtained from integrating the heat source.  

\section{\label{sec:C}Conclusion}

We presented a predictive description of the plasma sheath that evaluates plasma profiles throughout the quasineutral plasma region up to the wall using a hierarchy of fluid equations. The employed PINN approach was shown to provide a robust means of treating sheath models incorporating a diverse range of physics including finite ion temperature, ion-neutral collisions, collisional ionization, and heat transport. While the offline training time of the PINN was long compared to a traditional fluid solver, the online prediction time is far faster than a typical solver, enabling sheath profiles to be quickly inferred. Further noting that a single PINN can be trained across a range of physics parameters, the resulting PINN can be used as a rapid surrogate for inferring how distinct physical parameters impact the structure of the plasma sheath. Aside from providing insights into the sheath structure, we anticipate a future extension of this approach can be adapted to provide a robust tool for computing sheath boundary conditions to be used by quasilinear plasma codes. 

We began by identifying a parametric solution to the classic sheath problem with a PINN, where the predictions of the PINN were based purely on physical constraints without incorporating any simulation or experimental data. The profiles predicted by the PINN together with sheath properties such as the potential drop, sheath edge to density ratio, ion speed at the sheath entrance, and sheath width were compared with solutions obtained from a Runge-Kutta solver, where we observed excellent agreement across a range of physics parameters. This model was then extended to include a self-consistent collisional ionization source, where it was found that ionization can substantially modify the sheath solution at substantial electron temperatures and high electron and neutral densities. In the final model, we included an electron temperature equation, allowing for the electron temperature profile to be computed self-consistently. The present study assumed the collisional limit and closed the fluid system with Braginskii's expressions \cite{braginskii1965transport} for the heat flux, momentum exchange and energy exchange terms. No free parameters were considered in this work, though we do not anticipate extending this model to incorporate a parametric solution to pose a substantial hurdle. 

Finally, while the present work considers fluid models of the plasma sheath, by developing sheath models that are capable of solving sheath profiles for a diverse range of physical parameters and closures, this provides the first step toward setting up an inverse problem for determining more accurate closures for the plasma sheath. In this approach, data from a particle-in-cell (PIC)~\cite{li2022bohm} or continuum kinetic code~\cite{cagas2017continuum} would be used to inform a more accurate closure of the fluid system, thus enabling an efficient, though high-physics fidelity description of the sheath to be achieved. This last thrust will be left for future work, though we anticipate the fluid models developed in this work to provide the first step toward identifying this more accurate closure.

\begin{acknowledgements}

This work was supported by the Department of Energy, Office of Fusion Energy Sciences under award DE-SC0024634 and the National Science Foundation under award 2511564. The authors acknowledge the University of Florida Research Computing for providing computational resources that have contributed to the research results reported in this publication. 

\end{acknowledgements}


\appendix

\section{Runge-Kutta Solution to Sheath Equations}

This appendix briefly describes the Runge-Kutta solver used to verify the PINN solution.

\subsection{\label{sec:AppConstant}Constant Source}
In order to verify the PINN solution, Eqs. (\ref{eq:218}) and (\ref{eq:219}) can be written as ODEs and solved by a numerical scheme such as Runga-Kutta. To put the sheath equations in a form that they can be readily integrated, Poisson's equation is split into two ODEs:
\begin{equation}
    \label{eq:21NM1} 
    \frac{\partial \phi}{\partial x} = -E,
\end{equation}
\begin{equation}
    \label{eq:21NM2} 
    \frac{\partial E}{\partial x} = \frac{x}{Lu_i}-n_we^\phi,
\end{equation}
where from Eq. (\ref{eq:216}) we can write $n_i = x/(Lu_i)$ and from $\Gamma_w=1$ the density of the electrons at the wall can be written $n_w=1/u_{e,wall}$, with the electron velocity at the wall given by $u_{e,wall}=\sqrt{m_i/2\pi m_e}$. Additionally, the ion momentum equation [Eq. (\ref{eq:219})] becomes:
\begin{equation}
    \label{eq:21NM3} 
    \frac{\partial u_i}{\partial x} = \frac{E}{u_i}-\frac{u_i}{x}-\left(\frac{\lambda_{De}}{\lambda_{in}}\right),
\end{equation}
where we have assumed that $T_i=0$. We immediately see that a singularity exists at $x=0$ and $u_i=0$, such that the ODE integration cannot begin at $x=0$. Furthermore, the system is not automatically bounded in terms of where it should finish. These issues can be solved by starting the simulation a small distance from $x=0$ and analytically calculating a potential drop to indicate where the wall is located. We instead made predictions from the PINN at $x=0.01$ for the density, velocity, and potential, setting the initial conditions to agree with the PINN solution. Then, the RK simulation ended once the potential fell to zero. When using the numerical solver in this way, it checks that the PINN found a solution to the equations.  

\subsection{\label{sec:AppComplex}Complex Source}
Similar to the simple source case, this system, with the more complicated source, can be written into a series of ODEs, provided the ion temperature is assumed to be 0. The system is as follows:

\begin{equation}
    \label{eq:22NM1} 
    \frac{\partial \phi}{\partial x} = -E,
\end{equation}

\begin{equation}
    \label{eq:22NM2} 
    \frac{\partial E}{\partial x} = n_we^{\phi}\left(\frac{u_e}{u_i}-1\right),
\end{equation}

\begin{equation}
    \label{eq:22NM3} 
    \frac{\partial u_i}{\partial x} = \frac{E}{u_i} - \frac{u_iS}{u_en_we^{\phi}} - n_n\sigma_{in},
\end{equation}

\begin{equation}
    \label{eq:22NM4} 
    \frac{\partial u_e}{\partial x} = \frac{S}{n_we^{\phi}} + u_eE,
\end{equation}
where $S=1/L + n_nn_we^{\phi}\bar{S}_{ionize}$
, where $\bar{S}_{ionize}$ is defined above--below Eq. (\ref{eq:2210}). Dealing with finite ion temperature is difficult for numerical methods, thus we will only consider the limiting case of $T_i=0$. The model suffers from singularities, like the previous model, and the same strategies were employed to mitigate them. For a detailed derivation of the ion-neutral cross-section, see the appendix of reference \cite{beving2022sheath}, which is based on \cite{baalrud2012transport}.

\bibliographystyle{apsrev}
\bibliography{References}

@preamble{ " \newcommand{\noop}[1]{} " }

@preamble{"\def\authornoop#1{}"}

@article{wang2021understanding,
  title={Understanding and mitigating gradient flow pathologies in physics-informed neural networks},
  author={Wang, Sifan and Teng, Yujun and Perdikaris, Paris},
  journal={SIAM Journal on Scientific Computing},
  volume={43},
  number={5},
  pages={A3055--A3081},
  year={2021},
  publisher={SIAM}
}

@article{ahn2025deep,
  title={Deep transfer operator learning for predicting low temperature plasma sheath dynamics in semiconductor processing},
  author={Ahn, Sangjun and Bae, Jinkyu and Yoo, Suyoung and Nam, Sang Ki},
  journal={Physics of Plasmas},
  volume={32},
  number={9},
  pages={093505},
  year={2025},
  publisher={AIP Publishing}
}

@article{jang2024grad,
  title={Grad--Shafranov equilibria via data-free physics informed neural networks},
  author={Jang, Byoungchan and Kaptanoglu, Alan A and Gaur, Rahul and Pan, Shaowu and Landreman, Matt and Dorland, William},
  journal={Physics of Plasmas},
  volume={31},
  number={3},
  pages={032510},
  year={2024},
  publisher={AIP Publishing}
}

@article{cagas2017continuum,
  title={Continuum kinetic and multi-fluid simulations of classical sheaths},
  author={Cagas, Petr and Hakim, Ammar and Juno, James and Srinivasan, Bhuvana},
  journal={Physics of Plasmas},
  volume={24},
  number={2},
  pages={022118},
  year={2017},
  publisher={AIP Publishing}
}

@article{osborne2026physics,
  title={Physics-informed neural network for rapid ion trajectory prediction in ion-etch reactors},
  author={Osborne, Garrison and Roy, Sarthak and Webb, Ethan and McDevitt, Christopher J and Roy, Subrata},
  journal={Physics of Fluids},
  volume={38},
  number={3},
  pages={037119},
  year={2026},
  publisher={AIP Publishing}
}

@article{burby2020fast,
  title={Fast neural Poincar{\'e} maps for toroidal magnetic fields},
  author={Burby, Joshua William and Tang, Qi and Maulik, Romit},
  journal={Plasma Physics and Controlled Fusion},
  volume={63},
  number={2},
  pages={024001},
  year={2020},
  publisher={IOP Publishing}
}

@article{toscano2025pinns,
  title={From pinns to pikans: Recent advances in physics-informed machine learning},
  author={Toscano, Juan Diego and Oommen, Vivek and Varghese, Alan John and Zou, Zongren and Ahmadi Daryakenari, Nazanin and Wu, Chenxi and Karniadakis, George Em},
  journal={Machine Learning for Computational Science and Engineering},
  volume={1},
  number={1},
  pages={1--43},
  year={2025},
  publisher={Springer}
}

@article{wang2023expert,
  title={An expert's guide to training physics-informed neural networks},
  author={Wang, Sifan and Sankaran, Shyam and Wang, Hanwen and Perdikaris, Paris},
  journal={arXiv preprint arXiv:2308.08468},
  year={2023}
}

@article{mathews2022deep,
  title={Deep Electric Field Predictions by Drift-Reduced Braginskii Theory with Plasma-Neutral Interactions Based on Experimental Images of Boundary Turbulence},
  author={Mathews, Abhilash and Hughes, Jerry W and Terry, James L and Baek, Seung-Gyou},
  journal={Physical Review Letters},
  volume={129},
  number={23},
  pages={235002},
  year={2022},
  publisher={APS}
}

@article{karniadakis2021physics,
  title={Physics-informed machine learning},
  author={Karniadakis, George Em and Kevrekidis, Ioannis G and Lu, Lu and Perdikaris, Paris and Wang, Sifan and Yang, Liu},
  journal={Nature Reviews Physics},
  volume={3},
  number={6},
  pages={422--440},
  year={2021},
  publisher={Nature Publishing Group}
}

@article{chodura1982plasma,
  title={Plasma--wall transition in an oblique magnetic field},
  author={Chodura, Roland},
  journal={The Physics of Fluids},
  volume={25},
  number={9},
  pages={1628--1633},
  year={1982},
  publisher={American Institute of Physics}
}

@article{beving2022sheath,
  title={How sheath properties change with gas pressure: modeling and simulation},
  author={Beving, LP and Hopkins, MM and Baalrud, SD},
  journal={Plasma Sources Science and Technology},
  volume={31},
  number={8},
  pages={084009},
  year={2022},
  publisher={IOP Publishing}
}

@article{mcdevitt2024physics,
  title={A physics-informed deep learning description of Knudsen layer reactivity reduction},
  author={McDevitt, Christopher J and Tang, Xian-Zhu},
  journal={Physics of Plasmas},
  volume={31},
  number={6},
  pages={062701},
  year={2024},
  publisher={AIP Publishing}
}

@book{stangeby,
  title={The plasma boundary of magnetic fusion devices},
  author={Stangeby, Peter C and others},
  volume={224},
  year={2000},
  publisher={Institute of Physics Pub. Philadelphia, Pennsylvania}
}

@article{robertson2013sheaths,
  title={Sheaths in laboratory and space plasmas},
  author={Robertson, Scott},
  journal={Plasma Physics and Controlled Fusion},
  volume={55},
  number={9},
  pages={093001},
  year={2013},
  publisher={IOP Publishing}
}

@article{Raissi,
  title={Physics-informed neural networks: A deep learning framework for solving forward and inverse problems involving nonlinear partial differential equations},
  author={M. Raissi and P. Perdikaris and G.E. Karniadakis},
  journal={Journal of Computational Physics},
  volume={378},
  pages={686--707},
  year={2019},
  publisher={Elsevier Inc.}
}

@article{trieschmann2023machine,
  title={Machine learning for advancing low-temperature plasma modeling and simulation},
  author={Trieschmann, Jan and Vialetto, Luca and Gergs, Tobias},
  journal={Journal of Micro/Nanopatterning, Materials, and Metrology},
  volume={22},
  number={4},
  pages={041504--041504},
  year={2023},
  publisher={Society of Photo-Optical Instrumentation Engineers}
}

@article{godyak1982,
  title={Modified Bohm criterion for a collisional plasma},
  author={Godyak, Valery A},
  journal={Physics Letters A},
  volume={89},
  number={2},
  pages={80--81},
  year={1982},
  publisher={Elsevier}
}

@article{riemann1997collisions,
  title={The influence of collisions on the plasma sheath transition},
  author={Riemann, K-U},
  journal={Physics of plasmas},
  volume={4},
  number={11},
  pages={4158--4166},
  year={1997},
  publisher={American Institute of Physics}
}

@article{raimbault2008edge,
  title={Edge-to-center plasma density ratio in high density plasma sources},
  author={Raimbault, Jean-Luc and Chabert, Pascal},
  journal={Plasma Sources Science and Technology},
  volume={18},
  number={1},
  pages={014017},
  year={2008},
  publisher={IOP Publishing}
}

@article{li2022bohm,
  title={Bohm criterion of plasma sheaths away from asymptotic limits},
  author={Li, Yuzhi and Srinivasan, Bhuvana and Zhang, Yanzeng and Tang, Xian-Zhu},
  journal={Physical Review Letters},
  volume={128},
  number={8},
  pages={085002},
  year={2022},
  publisher={APS}
}

@article{cuomo2022,
  title={Scientific machine learning through physics--informed neural networks: Where we are and what’s next},
  author={Cuomo, Salvatore and Di Cola, Vincenzo Schiano and Giampaolo, Fabio and Rozza, Gianluigi and Raissi, Maziar and Piccialli, Francesco},
  journal={Journal of Scientific Computing},
  volume={92},
  number={3},
  pages={88},
  year={2022},
  publisher={Springer}
}

@article{brunton2016,
  title={Discovering governing equations from data by sparse identification of nonlinear dynamical systems},
  author={Brunton, Steven L and Proctor, Joshua L and Kutz, J Nathan},
  journal={Proceedings of the national academy of sciences},
  volume={113},
  number={15},
  pages={3932--3937},
  year={2016},
  publisher={National Acad Sciences}
}

@article{rudy2017,
  title={Data-driven discovery of partial differential equations},
  author={Rudy, Samuel H and Brunton, Steven L and Proctor, Joshua L and Kutz, J Nathan},
  journal={Science advances},
  volume={3},
  number={4},
  pages={e1602614},
  year={2017},
  publisher={American Association for the Advancement of Science}
}

@article{alves2022,
  title={Data-driven discovery of reduced plasma physics models from fully kinetic simulations},
  author={Alves, E Paulo and Fiuza, Frederico},
  journal={Physical Review Research},
  volume={4},
  number={3},
  pages={033192},
  year={2022},
  publisher={APS}
}

@article{alvarez2020plasma,
  title={Plasma-sheath transition in multi-fluid models with inertial terms under low pressure conditions: Comparison with the classical and kinetic theory},
  author={Alvarez-Laguna, Alejandro and Magin, Thierry and Massot, Marc and Bourdon, Anne and Chabert, Pascal},
  journal={Plasma Sources Science and Technology},
  volume={29},
  number={2},
  pages={025003},
  year={2020},
  publisher={IOP Publishing}
}

@article{crespo2018positive,
  title={Positive ion temperature effect on the plasma-wall transition},
  author={Morales Crespo, R},
  journal={Physics of Plasmas},
  volume={25},
  number={6},
  pages={063509},
  year={2018},
  publisher={AIP Publishing}
}

@article{kaganovich2020physics,
  title={Physics of E$\times$ B discharges relevant to plasma propulsion and similar technologies},
  author={Kaganovich, Igor D and Smolyakov, Andrei and Raitses, Yevgeny and Ahedo, Eduardo and Mikellides, Ioannis G and Jorns, Benjamin and Taccogna, Francesco and Gueroult, Renaud and Tsikata, Sedina and Bourdon, Anne and others},
  journal={Physics of Plasmas},
  volume={27},
  number={12},
  pages={120601},
  year={2020},
  publisher={AIP Publishing}
}

@article{lieberman1994principles,
  title={Principles of plasma discharges and materials processing},
  author={Lieberman, Michael A and Lichtenberg, Allan J},
  journal={MRS Bulletin},
  volume={30},
  number={12},
  pages={899--901},
  year={1994}
}

@article{hutchinson2002principles,
  title={Principles of plasma diagnostics},
  author={Hutchinson, Ian H},
  journal={Plasma Physics and Controlled Fusion},
  volume={44},
  number={12},
  pages={2603--2603},
  year={2002}
}

@article{hershkowitz2005sheaths,
  title={Sheaths: More complicated than you think},
  author={Hershkowitz, Noah},
  journal={Physics of plasmas},
  volume={12},
  number={5},
  pages={055502},
  year={2005},
  publisher={AIP Publishing}
}

@article{baalrud2009instability,
  title={Instability-Enhanced Collisional Friction Can Determine the Bohm Criterion<? format?> in Multiple-Ion-Species Plasmas},
  author={Baalrud, SD and Hegna, CC and Callen, JD},
  journal={Physical review letters},
  volume={103},
  number={20},
  pages={205002},
  year={2009},
  publisher={APS}
}

@inproceedings{mcdevitt2024Navier,
  title={Physics-constrained deep learning of incompressible cavity flows},
  author={McDevitt, Christopher and Fowler, Eric and Roy, Subrata},
  booktitle={AIAA Scitech 2024 Forum},
  pages={1692},
  year={2024}
}

@inproceedings{abadi2016tensorflow,
  title={$\{$TensorFlow$\}$: a system for $\{$Large-Scale$\}$ machine learning},
  author={Abadi, Mart{\'\i}n and Barham, Paul and Chen, Jianmin and Chen, Zhifeng and Davis, Andy and Dean, Jeffrey and Devin, Matthieu and Ghemawat, Sanjay and Irving, Geoffrey and Isard, Michael and others},
  booktitle={12th USENIX symposium on operating systems design and implementation (OSDI 16)},
  pages={265--283},
  year={2016}
}

@article{kiyani2025optimizer,
  title={Which Optimizer Works Best for Physics-Informed Neural Networks and Kolmogorov-Arnold Networks?},
  author={Kiyani, Elham and Shukla, Khemraj and Urb{\'a}n, Jorge F and Darbon, J{\'e}r{\^o}me and Karniadakis, George Em},
  journal={arXiv e-prints},
  pages={arXiv--2501},
  year={2025}
}

@article{urban2025unveiling,
  title={Unveiling the optimization process of Physics Informed Neural Networks: How accurate and competitive can PINNs be?},
  author={Urb{\'a}n, Jorge F and Stefanou, Petros and Pons, Jos{\'e} A},
  journal={Journal of Computational Physics},
  volume={523},
  pages={113656},
  year={2025},
  publisher={Elsevier}
}

@article{braginskii1965transport,
  title={Transport processes in a plasma},
  author={Braginskii, SI},
  journal={Reviews of plasma physics},
  volume={1},
  pages={205},
  year={1965}
}

@article{riemann2005plasma,
  title={The plasma--sheath matching problem},
  author={Riemann, Karl U and Seebacher, Josef and Tskhakaya, DD and Kuhn, Siegbert},
  journal={Plasma physics and controlled fusion},
  volume={47},
  number={11},
  pages={1949},
  year={2005},
  publisher={IOP Publishing}
}

@book{helander2005collisional,
  title={Collisional transport in magnetized plasmas},
  author={Helander, Per and Sigmar, Dieter J},
  volume={4},
  year={2005},
  publisher={Cambridge university press}
}

@article{richardson20192019,
  title={2019 NRL plasma formulary},
  author={Richardson, Andrew S},
  year={2019}
}

@article{sun2020surrogate,
  title={Surrogate modeling for fluid flows based on physics-constrained deep learning without simulation data},
  author={Sun, Luning and Gao, Han and Pan, Shaowu and Wang, Jian-Xun},
  journal={Computer Methods in Applied Mechanics and Engineering},
  volume={361},
  pages={112732},
  year={2020},
  publisher={Elsevier}
}

@book{bishop2006pattern,
  title={Pattern recognition and machine learning},
  author={Bishop, Christopher M and Nasrabadi, Nasser M},
  volume={4},
  number={4},
  year={2006},
  publisher={Springer}
}

@article{lu2021deepxde,
  author  = {Lu, Lu and Meng, Xuhui and Mao, Zhiping and Karniadakis, George Em},
  title   = {{DeepXDE}: A deep learning library for solving differential equations},
  journal = {SIAM Review},
  volume  = {63},
  number  = {1},
  pages   = {208-228},
  year    = {2021},
  doi     = {10.1137/19M1274067}
}

@article{kingma2014adam,
  title={Adam: A method for stochastic optimization},
  author={Kingma, Diederik P},
  journal={arXiv preprint arXiv:1412.6980},
  year={2014}
}

@article{liu1989limited,
  title={On the limited memory BFGS method for large scale optimization},
  author={Liu, Dong C and Nocedal, Jorge},
  journal={Mathematical programming},
  volume={45},
  number={1},
  pages={503--528},
  year={1989},
  publisher={Springer}
}

@article{colella1999conservative,
  title={A conservative finite difference method for the numerical solution of plasma fluid equations},
  author={Colella, Phillip and Dorr, Milo R and Wake, Daniel D},
  journal={Journal of Computational Physics},
  volume={149},
  number={1},
  pages={168--193},
  year={1999},
  publisher={Elsevier}
}

@article{van1995neural,
  title={Neural network differential equation and plasma equilibrium solver},
  author={van Milligen, B Ph and Tribaldos, V and Jim{\'e}nez, JA},
  journal={Physical review letters},
  volume={75},
  number={20},
  pages={3594},
  year={1995},
  publisher={APS}
}

@book{holmes2007introduction,
  title={Introduction to numerical methods in differential equations},
  author={Holmes, Mark H},
  year={2007},
  publisher={Springer}
}

@misc{Phelps,
    title = {{Phelps database}},
    howpublished = {\url{www.lxcat.net/Phelps}},
    note = {Accessed: 6 December 2025}
}

@article{baalrud2012transport,
  title={Transport coefficients in strongly coupled plasmas},
  author={Baalrud, Scott D},
  journal={Physics of Plasmas},
  volume={19},
  number={3},
  pages={030701},
  year={2012},
  publisher={AIP Publishing}
}

@article{paszke2019pytorch,
  title={Pytorch: An imperative style, high-performance deep learning library},
  author={Paszke, Adam and Gross, Sam and Massa, Francisco and Lerer, Adam and Bradbury, James and Chanan, Gregory and Killeen, Trevor and Lin, Zeming and Gimelshein, Natalia and Antiga, Luca and others},
  journal={Advances in neural information processing systems},
  volume={32},
  year={2019}
}

@article{sternberg1996approximation,
  title={Approximation of the bounded plasma problem by the plasma and the sheath models},
  author={Sternberg, Natalia and Godyak, Valery A},
  journal={Physica D: Nonlinear Phenomena},
  volume={97},
  number={4},
  pages={498--508},
  year={1996},
  publisher={Elsevier}
}

@article{valentini1988removal,
  title={Removal of singularities in the hydrodynamic description of plasmas including space-charge effects, several species of ions and non-vanishing ion temperature},
  author={Valentini, H-B},
  journal={Journal of Physics D: Applied Physics},
  volume={21},
  number={2},
  pages={311--321},
  year={1988}
}

\end{document}